\documentstyle[aps,multicol,epsf]{revtex}
\tighten
\begin{document}
\draft

\title{Local densities, distribution functions, and wave function 
correlations for spatially resolved shot noise at nanocontacts}

\author{Thomas Gramespacher and Markus B\"uttiker}
\address{D\'epartement de Physique Th\'eorique, Universit\'e 
de Gen\`eve, CH-1211, Gen\`eve 4, Switzerland} 
\date{\today}
\maketitle

\begin{abstract}
We consider a current-carrying, phase-coherent multi-probe conductor to which
a small tunneling contact is attached. We treat the conductor and the tunneling
contact as a phase-coherent entity and use a Green's function formulation
of the scattering approach. We show that the average current and the current
fluctuations at the tunneling contact are determined by an {\em effective local
non-equilibrium distribution function}. This function characterizes the
distribution of charge-carriers (or quasi-particles) inside the conductor.
It is an exact quantum-mechanical expression and contains the phase-coherence
of the particles via local partial densities of states, called
{\em injectivities}. The distribution function is analyzed for different
systems in the zero-temperature limit as well as at finite temperature.
Furthermore, we investigate in detail the correlations of the currents measured
at two different contacts of a four-probe sample, where two of the probes
are only weakly coupled contacts. In particular, we show that the correlations
of the currents are at zero-temperature given by spatially {\em non-diagonal}
injectivities and emissivities. These non-diagonal densities are sensitive to 
correlations of wave functions
and the phase of the wave functions. We consider ballistic conductors
and metallic diffusive conductors. We also analyze the Aharonov-Bohm  
oscillations in the shot noise correlations of a conductor which in the 
absence of the nano-contacts exhibits no flux-sensitivity in the conductance. 
\end{abstract}


\begin{multicols}{2}

\narrowtext

\section{Introduction}

To measure the properties of a system it is necessary 
to couple a measurement apparatus to the system. 
To minimize the effect that the presence of the
measurement apparatus has on the properties of the system, it is desirable 
to have the coupling as weak as possible. We are interested in the properties
of a current carrying, phase-coherent multi-probe conductor. 
Weak coupling or
non-invasive contacts on mesoscopic conductors
were already used by Engquist and Anderson\cite{engquist}
to re-derive Landauer's
resistance formula\cite{landauerres} for a small conductor with a scatterer.
Here we are interested in weak coupling contacts 
which are sensitive to the phase of current 
amplitudes\cite{buttiker86,buttiker89,gramespacher97}
and not only as in the work of Engquist and Anderson 
and related work\cite{imry} to absolute values of currents.  
Nowadays, the scanning tunneling microscope (STM)\cite{binnig82}
is a very powerful
experimental realization of a weakly coupled contact. Due to the fact that
the tunneling current to the tip originates only from an atomically small 
area on the surface below the tip, it has become the standard tool to 
measure the local electronic structure on the surface of conductors. 
In experiment, it is
possible to map the topography of a surface with atomic
resolution\cite{wiesendanger,avouris95}.
Standing electron wave patterns confined to quantum
corrals\cite{crommie}, which were constructed by manipulation of single atoms,
or on carbon nanotubes serving as a one-dimensional electron box\cite{dekker}
are clearly visible using a low-temperature STM.

In the theoretical description, initially
Tersoff and Hamann\cite{tersoff} used the Bardeen approach to
tunneling\cite{bardeen} to relate the tunneling conductance to the local
density of states (LDOS) $\nu(x)$
on the surface of the conductor. 
Recently, Bracher {\em et al.}\cite{bracher}
arrived at the same result using a propagator theory where the tip was
described as a localized source or sink of electrons. 
These approaches have
been used to
interpret many of the features encountered in STM images. 
In theory and
experiment, the STM has most often been used in a two-terminal configuration,
the two terminals being the tip on one side and the conductor on the other. 
The
current at the tip is then determined by the two-terminal conductance between
tip and surface, and is given by the Bardeen formula\cite{bardeen}, 
$G_{ts}=(e^2/h)4\pi^2\nu_{tip}|t|^2\nu(x)$, with the LDOS $\nu(x)$ of the
sample, $\nu_{tip}$ of the tunneling tip, and the coupling energy $|t|$.
The zero-bias
conductance thus measures directly the LDOS $\nu(x)$ on the surface of the 
conductor
below the tip.
\begin{figure}
\epsfxsize8cm
\epsffile{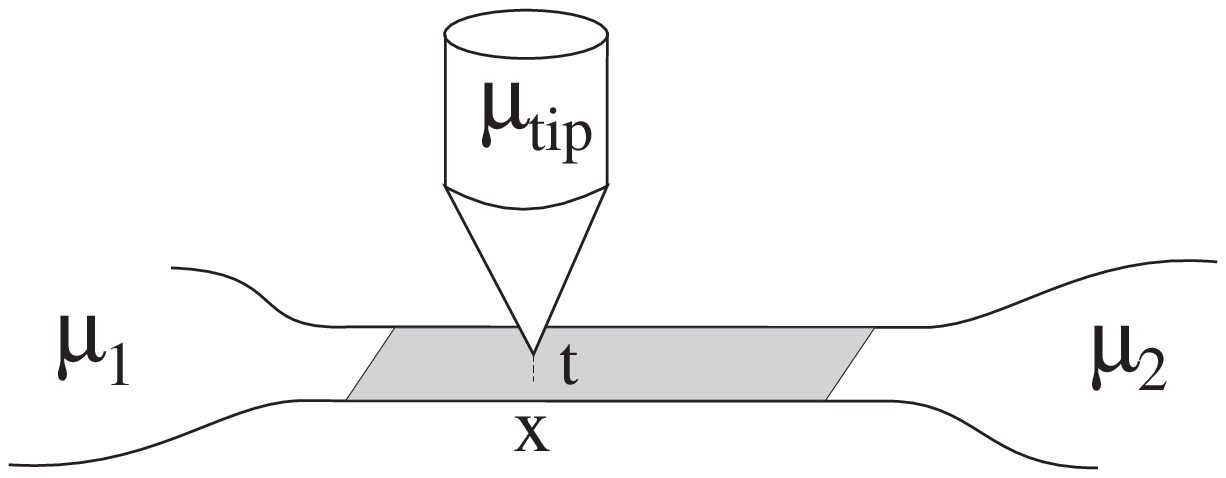}
\vspace{0.3cm}
\caption{Experimental setup to measure the effective local distribution
function. The tip of an STM couples at a point $x$ with a coupling strength $t$
to the surface of a multi-terminal conductor. The contacts of the conductor are
held at potentials $\mu_\alpha$ and the tip at potential $\mu_{tip}$.
This configuration
can be used to measure the time dependent current or voltage at the tip.}
\label{eintip}
\end{figure}

\begin{figure}
\epsfxsize8cm
\epsffile{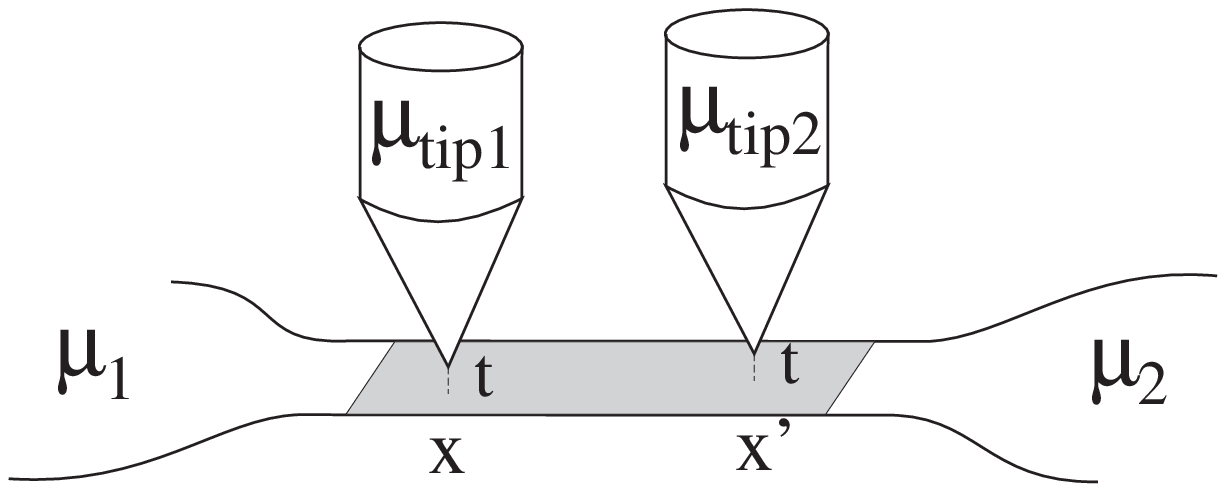}
\vspace{0.3cm}
\caption{Experimental setup to measure current correlations. Two STM tips are
coupled with strength $t$ at the positions $x$ and $x^\prime$ to the surface
of a small wire. The shaded region can be a metallic diffusive or a perfect
ballistic wire.}
\label{zweitips}
\end{figure}

In this article, we make theoretical predictions for measurements using
one (or two) tunneling tips on mesoscopic phase-coherent multi-probe
conductors and analyze the voltage and the current fluctuations 
measured at such a contact. 
The proposed experimental setup is shown in Figs.\ \ref{eintip}
and \ref{zweitips}.
The current at the tips is now determined by all conductances between the
tip and the massive contacts of the sample.
Applying a bias at the massive contacts of the multi-probe
conductor one can drive already a current through it without the presence of
the tunneling contacts. This puts the conductor into a
non-equilibrium state. Here we are interested in the characterization of 
the transport state. The STM is used to measure the electronic structure
on the surface of the current carrying sample. We will see later that the
average current and the current fluctuation spectrum at a single tunneling tip
are determined by an 
{\em effective local non-equilibrium distribution function}
expressed as a function of local partial densities of states (LPDOS) and the
Fermi distribution functions in the electron reservoirs.
A measurement of such a distribution,
averaged over a spatially wide area, 
has been performed by Pothier {\em et al.}\cite{pothier}
using a large superconducting
tunneling contact on a metallic diffusive wire.
We are interested in characterizing the transport state not only
locally but also by its spatial and temporal correlations. 
The measurement of the correlations of the currents at {\it two} 
different tunneling tips 
is related to spatially non-diagonal densities of states and can
give information about correlations of wave functions and
the phase of the wave functions.

Here, we use a fully phase-coherent theory of weak
coupling contacts starting from the overall scattering matrix 
which includes the conductor and the tunneling contacts as 
one entity. 
A fully phase-coherent discussion of four-probe resistances 
measured with weakly coupled contacts has been presented in 
Ref. \ \cite{buttiker89}. In this work and in recent work by
the two of us\cite{gramespacher97}  such an approach has been used 
to investigate 
the local voltage measurements and phase-coherent resistance measurements
on mesoscopic wires. Of particular interest is the relationship 
of the transmission probabilities to densities of states 
which characterize the conductor. In the transport problem 
of interest here it is shown\cite{gramespacher97} 
that the densities of states which appear 
are partial densities of states, called {\it injectivity} for transmission
from a contact of the conductor into the tip, and called {\it emissivity}
for the transition from the tip into one of the contacts of the conductor.
The transmission probabilities from the sample into the tip 
and from the tip into the sample can be viewed as a generalization
of the well known Bardeen expression for the two-terminal weak coupling
contact\cite{gramespacher97}. The generalized densities of states, 
the injectivity and emissivity play a fundamental role also 
in the dynamic conductance of mesoscopic systems\cite{buttikerjphys,christen} 
and in the non-linear conductance of mesoscopic systems\cite{christen96,ma98}. 

In the following, we use the same approach but extend the
discussion to treat temporal  
current and voltage fluctuations and investigate the correlations of currents
measured at two tunneling contacts.
As we will show later, these measurements can reveal more
information about the electronic structure than can be found by pure 
conductance
measurements.
At elevated temperature and with an applied bias the fluctuations of the
currents can be devided into two contributions:
the thermal noise, which is proportional
to the temperature and an excess noise, called shot-noise, which is only
present when the system is biased\cite{buttiker92a,BdJ}.
The thermal noise is via a
fluctuation-dissipation theorem related to a conductance and 
does therefore not
contain more information about the conductor
than can be drawn from measuring conductances. 
However, the shot-noise, which is at zero-temperature the only source of
fluctuations, can give more information\cite{buttiker92a}. 
For instance, the shot-noise spectrum can be used to distinguish between different
conductance mechanisms, such as ballistic 
or diffusive conductance\cite{BdJ}.
The low-frequency
shot-noise spectrum has been used to identify the fractional charge of
the quasi-particles in the fractional quantum Hall regime\cite{glattli,picci},
and, recently,
van den Brom and van Ruitenbeek\cite{brom} used combined
conductance and shot-noise
measurements to determine the detailed mechanism of the electrical conductance
through atom-size metallic gold-contacts. Birk {\em et al.}\cite{birk95}
measured the shot noise
at an STM tip. 
Of particular interest are current-current 
cross-correlations\cite{buttiker90,buttiker92a,buttiker92b}
due to their sensitivity to the statistics of the carriers.
Specific predictions have been made for current correlations of conductors in 
high quantizing magnetic fields\cite{buttiker90,buttiker92a,buttiker92b},
for ballistic conductors \cite{thrl},
for metallic diffusive conductors with massive 
contacts\cite{blanter97,sukhorukov98}, for chaotic cavities \cite{langen97},
and for hybrid normal and superconducting systems \cite{anantram}.   
Very recently, measurements of current cross-correlations (the 
electric analog of the Hanburry Brown Twiss experiment)
have been reported by Henny {\em et al.}\cite{henny98} for a Hall bar 
geometry which permits 
the separation of incident and reflected 
carrier streams as suggested in Ref. \onlinecite{buttiker90} 
and by Oliver
{\em et al.}\cite{oliver}
for a conductor that exhibits probably elements both 
of ballistic electron motion and chaotic electron motion. 
More severe tests of our understanding of fluctuations arise from 
probing exchange effects in correlations due to the 
quantum mechanical indistinguishability of identical particles. 
We will discuss exchange effects below in some detail. Earlier discussions
of exchange effects in cross-correlations in mesoscopic conductors 
can be found in 
Refs.\ \cite{buttiker91,buttiker92a,buttiker92b,blanter97,sukhorukov98,langen97,anantram}. 
An experiment which investigates exchange effects 
in the noise at a single contact due to two incident carrier 
streams has been carried out by Liu et al. \cite{liu}. 
Therefore, we believe it to be
justified to assume, that the shot-noise measurements at local
tunneling contacts proposed in this work can in fact 
be done as well.

In order to be able to make statements about the local structure or
wave function correlations on the sample surface, the contact between tip and
sample should be local in the sense
that tunneling occurs
only over a region which is small compared to the variation of the LDOS
on the surface of the conductor. In general, this length scale is given by
the Fermi wavelength of the surface states.
Modern STM measurements show clearly that atomic resolution on metallic
surfaces can be achieved
using sufficiently sharp tips.
In addition, STM tips have the advantage that they can be moved around on a
surface so that it is possible to draw entire {\it maps} of e.\ g.\ the LDOS and
to study the spatial variation of the transport and noise properties.
The theory
we formulate below, however, is also valid for spatially fixed contacts
provided the contact is sufficiently small and in the regime of
tunneling. Especially, for the experiment with two tunneling contacts,
Fig.\ \ref{zweitips}, it might practically be much easier to use a setup
with one spatially fixed tunneling contact and one (movable) STM tip.

We note that some of the results presented below have already been published
in a shortened version in \cite{gramespacher98}.
\section{Hamiltonian formulation of the scattering matrix and the L(P)DOS}
We are concerned with open mesoscopic systems consisting of a finite part where
electrons are scattered and to which $N$ huge, macroscopic electron reservoirs are
attached. The phase coherence length for the electrons is supposed to be
much longer than
the spatial dimensions of the scattering region. Then, inelastic scattering
takes only place in the electron reservoirs. In each reservoir $\alpha$
the electrons are in
equilibrium and distributed according to a Fermi function characterized by the
electro-chemical potential $\mu_\alpha$ and the temperature $T_\alpha$.
The finite scattering region is described by a Hamiltonian $H_C$ and the
connection to the electron reservoirs is modeled by semi-infinite ideal
leads described by a Hamiltonian $H_L$. As a basis of the Hamiltonian $H_C$
we choose $M$ localized states $|x\rangle$ (where $M$ is a very big number),
\begin{equation}
H_C=\sum_{x,x^\prime}|x\rangle H_{xx^\prime}\langle x^\prime|\, .
\end{equation}
The Hilbert space of the semi-infinite leads is spanned by scattering states
$|\alpha m\rangle$
totally reflected at the boundary to the scattering region.
At an energy $E$ we have to sum over the scattering states of all open
channels in the leads, 
\begin{equation}
H_L=\sum_{\alpha=1}^N\sum_{m=1}^{N_\alpha}|\alpha m\rangle E\langle \alpha m|
\, .
\end{equation}
The index $\alpha$
gives the number of the reservoir and the index $m$ is the channel number
of the incoming electron. In reservoir $\alpha$ there are $N_\alpha$ open
channels at the energy $E$.
Finally, we have to describe the coupling between the scattering states in
the ideal leads and the conductor by a coupling matrix
\begin{equation}
W=\sum_x\sum_{\alpha=1}^N\sum_{m=1}^{N_\alpha}
|x\rangle W_{x,\alpha m}\langle\alpha m|\, .
\end{equation}
The Hamiltonian of the entire system
then reads
\begin{equation}
{\cal H}=H_L+H_C+W+W^\dagger\, .
\end{equation}
The Green's function between two points $x$ and $x^\prime$ inside the scattering
region is then at the Fermi-energy $E_F$ given by\cite{iida90}
\begin{equation}
G(x,x^\prime)=\langle x|(E_F-H_C+i\pi WW^\dagger)^{-1}|x^\prime\rangle\, .
\end{equation}
The matrix elements of the scattering matrix $s_{\alpha m,\beta n}$, which
describes the scattering of an incoming particle in channel $n$ of contact
$\beta$ being scattered into channel $m$ of contact $\alpha$, can be 
written as
\begin{equation}
s_{\alpha m,\beta n}=\delta_{\alpha\beta}\delta_{mn}
-2\pi i\sum_{x,x^\prime}W_{x,\alpha m}^* G(x,x^\prime)W_{x^\prime,\beta n}
\, .\label{scatgreen}
\end{equation}
The scattering matrix depends on the electrostatic potential $U(x)$ in the
scattering region which is included in the Hamiltonian $H_C$.
This potential has in principle to be calculated self-consistently for the
system in equilibrium\cite{buttikerjphys}.
A small variation $\delta\mu_\alpha$ of the
electro-chemical potential in a reservoir $\alpha$ injects
then at a position
$x$ inside the conductor an additional charge \cite{note1} $q(x)=e\nu(x,\alpha)\delta
\mu_\alpha$. The proportionality factor $\nu(x,\alpha)$ is a LPDOS and is
called the injectivity of contact $\alpha$ at the point $x$. 
It can be expressed
with the help of the scattering matrix as\cite{christen}
\begin{equation}
\nu(x,\alpha)  =  \frac{-1}{2\pi i}\sum_\beta\mbox{\rm Tr}\left(
{\bf s}_{\beta\alpha}^\dagger
\frac{\delta
{\bf s}_{\beta\alpha}}{e\delta U(x)}\right) 
\end{equation}
Here, ${\bf s}_{\alpha\beta}$ denotes the
$N_\alpha\times N_\beta$ sub-matrix of the
scattering matrix which describes the scattering of electrons between all
channels of contacts $\alpha$ and $\beta$. 
With the Green's function defined above, we have for the injectivity
the expression\cite{gramespacher97}
\begin{equation}
\nu(x,\alpha)=\langle x| G \Gamma_\alpha G^\dagger| x\rangle,\label{inj} \\
\end{equation}
where we introduced the abbreviation
\begin{equation}
\Gamma_\alpha=\sum_{x,x^\prime}|x\rangle\langle x^\prime|\sum_{m=1}^{N_\alpha}
W_{x,\alpha m}W_{x^\prime,\alpha m}^*\, .
\end{equation}
Using the Lippmann-Schwinger equation $|\psi_{\alpha m}\rangle
=(1-GW)|\alpha m\rangle$ which relates the scattering state $|\psi_{\alpha m}
\rangle$ of the entire coupled system to the scattering states $|\alpha m
\rangle$ of the isolated leads\cite{iida90},
one can express the injectivity in terms of
the scattering wave functions
\begin{equation}
\nu(x,\alpha)=
\sum_{m=1}^{N_\alpha}\frac{1}{hv_{\alpha m}}|\psi_{\alpha m}(x)|^2
\label{injscatstat}
\end{equation}
Here, $v_{\alpha m}=\sqrt{2/m^\star(E_F-E_{\alpha m}^0)}$ is the velocity
of an incoming electron at the Fermi energy $E_F$
in channel $m$ of contact $\alpha$, $m^\star$ is the
effective electron mass and $E_{\alpha m}^0$ is the threshold energy of
channel $m$ of contact $\alpha$.

Related to the injectivity is another LPDOS, the emissivity $\nu(\beta,x)$
of a point $x$ into contact $\alpha$, defined as
\begin{equation}
\nu(\beta,x) = \frac{-1}{2\pi i}\sum_\alpha\mbox{\rm Tr}\left(
{\bf s}_{\beta\alpha}^\dagger
\frac{\delta
{\bf s}_{\beta\alpha}}{e\delta U(x)}\right)
\end{equation}
and in terms of Green's functions given by
\begin{equation}
\nu(\beta,x) = \langle x|G^\dagger\Gamma_\beta G|x\rangle\, .\label{em}
\end{equation}
If there
is a magnetic field $B$ present, the injectivity and emissivity obey
the symmetry\cite{christen}
\begin{equation}
\nu_B(\alpha,x)=\nu_{-B}(x,\alpha)\, .\label{injsym}
\end{equation}
That means, reversing the magnetic field turns the injectivity of a specific
contact
into its emissivity and vice versa.
As a special case, Eq.\ (\ref{injsym}) states that injectivity and
emissivity are the same if there is no magnetic field present.
Furthermore, the emissivity can according to Eq.\ (\ref{injscatstat})
and (\ref{injsym}) be expressed in terms of the scattering states of the
Hamiltonian with the reversed magnetic field.
The LDOS $\nu(x)$ is the sum of the injectivities of all
contacts or the emissivities of all contacts,
\begin{equation}
\nu(x)=\sum_\alpha \nu(x,\alpha)=\sum_\beta\nu(\beta,x)\, .
\end{equation}
The LDOS is therefore invariant under reversal of the magnetic field.

The form of Eqs.\ (\ref{inj}) and (\ref{em}) suggests to define a
non-diagonal two-point injectivity by
\begin{eqnarray}
\nu(x^\prime,x,\alpha) & = &  
\langle x^\prime |G\Gamma_\alpha G^\dagger|x \rangle
 \label{nondiaginj}\\
& = & \sum_{m=1}^{N_\alpha}\frac{1}{hv_{\alpha m}}
\psi_{\alpha m}(x^\prime)\psi_{\alpha m}(x)^*
\end{eqnarray}
and analogously a non-diagonal two-point emissivity by
\begin{equation}
\nu(\beta,x^\prime,x)=
\langle x^\prime |G^\dagger\Gamma_\beta G|x \rangle\, .\label{nondiagem}
\end{equation}
In fact, we will see that it is exactly these spatially non-diagonal LPDOS which
determine the correlation of the currents at two tips.
\section{Scattering matrix formulation of current and noise}
Our goal is to investigate the local electronic structure of
a mesoscopic phase-coherent multi-probe conductor using one or
several locally weakly coupled probes such as e.\ g.\ STM tips.
One can think of transport experiments which measure the average
current determined by conductances or one can measure the
fluctuations of the current away from its average.
The scattering matrix approach has proven to be very useful in describing
transport and noise measurements at multi-probe conductors\cite{dattabook}.
It provides us with formulae which express
the currents and the fluctuations of the currents at the contacts of a multi-probe
conductor in terms of its scattering matrix and the Fermi functions $f_\alpha
(T,E)$ of the electron distribution in the reservoirs.
The experimentally directly accessible parameters of the system are the
temperature $T$ and the electro-chemical potentials $\mu_\alpha$ in the
large electron reservoirs.

For a certain temperature $T$ and given potentials
the average current flowing from contact $\alpha$
into the conductor is\cite{buttiker92a}
\begin{equation}
\langle I_\alpha\rangle=\frac{e}{h}\sum_\beta\int dE
\mbox{\rm Tr}\left[ A_{\beta\beta}(\alpha)\right]
f_\beta(E) \label{cond}
\end{equation}
with the current matrix $A_{\delta\gamma}(\alpha)
={\bf 1}_\alpha\delta_{\alpha\delta}
\delta_{\alpha\gamma}
-{\bf s}_{\alpha\delta}^\dagger(E) {\bf s}_{\alpha\gamma}(E)$.
The energy dependent transmission probability
between two different contacts $\alpha$ and $\beta$
is $T_{\alpha\beta}=-\mbox{\rm Tr}[ A_{\beta\beta}(\alpha)]$.
In the limit of zero temperature
and if we assume that the differences of the applied potentials
are so small that the scattering matrix depends only very weakly on energy
in the energy interval of interest, formula (\ref{cond}) reduces to
\begin{equation}
\langle I_\alpha\rangle =\frac{e}{h}\sum_\beta
T_{\alpha\beta}(\mu_\alpha-\mu_\beta)
\, ,\label{condlin}
\end{equation}
where the transmission probabilities $T_{\alpha\beta}$
have to be evaluated at the Fermi energy.

The correlation spectrum $\langle\Delta I_\alpha\Delta I_\beta\rangle$ of
the currents at two contacts
$\alpha$ and $\beta$ is the Fourier transform of the current-current
correlator\cite{BdJ},
\begin{equation}
\langle\Delta I_\alpha\Delta I_\beta\rangle =\int dt e^{i\omega t}
\langle\Delta I_\alpha(t)\Delta I_\beta
(t+t_0)\rangle\, ,
\end{equation}
where $\Delta I_\alpha(t)= I_\alpha(t)-\langle I_\alpha \rangle$.
In the low-frequency limit, $\omega\rightarrow 0$, one gets\cite{buttiker92a}
\begin{eqnarray}
\langle\Delta I_\alpha\Delta I_\beta\rangle & = & \nonumber \\
 \frac{2e^2}{h}
& \sum_{\delta\gamma} &
\int dE \mbox{\rm Tr}
\left[ A_{\delta\gamma}(\alpha)A_{\gamma\delta}(\beta)\right]
f_\delta(1-f_\gamma)\, .\label{shotnoise}
\end{eqnarray}
For $\alpha=\beta$ this expression gives the low-frequency fluctuation
spectrum of the current at the contact $\alpha$. For $\alpha\neq\beta$
it gives the correlation-spectrum of the currents in the two contacts
$\alpha$ and $\beta$.
In general, the current fluctuation- or correlation-spectrum
is a mixture of thermal noise and, if the system
is biased, an excess noise called shot-noise.
At zero temperature all fluctuations in the currents are due to the
discreteness of the charge carriers. We are dealing with pure shot-noise. At a given
instant in time a carrier either arrives at a reservoir, i.\ e.\ a current
is measured, or it does not. Successive carriers that are totally uncorrelated
give the full (Poissonian) shot-noise, $S_{Poiss}=2e|\langle I\rangle |$. If
successive carriers are correlated, as is the case for electrons due to
Fermi statistics, the noise can be suppressed below this
value.

Eq.\ (\ref{shotnoise}) gives the fluctuation spectrum of the time-dependent
currents in the contacts under the condition that the potentials at the
reservoirs are held fixed and do not fluctuate. This corresponds to the case
where currents are measured using a zero-impedance external circuit. Alternatively,
we could measure the voltages at the reservoirs using ideal, infinite impedance
voltmeters. The infinite impedance external circuit then forces the currents
to be zero at all times, $I(t)=0$. Fluctuations in the currents have therefore
to be counterbalanced by fluctuations of the chemical potentials in the
electron reservoirs. In
linear response to the applied bias, current and potential are related
by a conductance matrix $G_{\alpha\beta}$,
\begin{equation}
I_\alpha(t)=\sum_\beta G_{\alpha\beta}(V_\beta+\Delta V_\beta(t))+\Delta
I_\alpha(t)\, , \label{volfluc1}
\end{equation}
where the $\Delta I_\alpha(t)$ are now considered as Langevin forces obeying the
correlation spectra given in Eq.\ (\ref{shotnoise}) and where we allowed the
potential at the reservoirs to be time dependent.
\begin{figure}
\epsfxsize6cm
\centerline{\epsffile{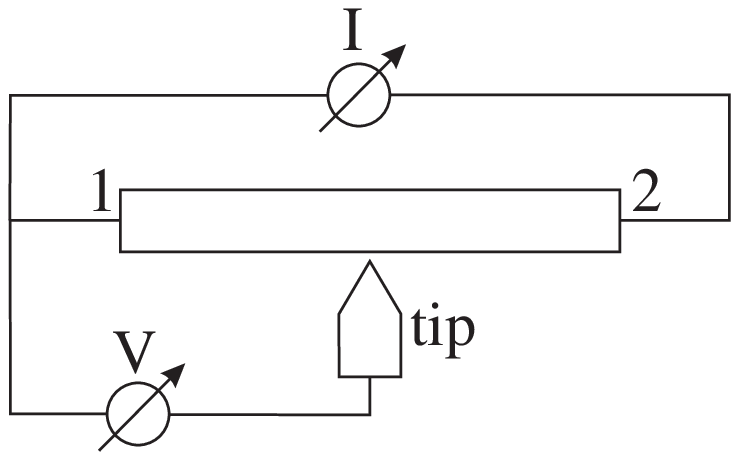}}
\vspace{0.3cm}
\caption{Experimental setup to measure the voltage fluctuations at the tip. The
voltage is measured using an infinite impedance voltmeter between contact 1 and
the tip, and the current is measured using a zero-impedance ampere-meter between
contacts 1 and 2.}
\label{volflucfig}
\end{figure}

Let us now consider the experimental
setup of Fig.\ \ref{volflucfig}.
We are interested in the fluctuations of the
voltage at the tip $\langle (\Delta V_{tip})^2\rangle$
measured relative to the voltage at contact 1. The current at the tip is always zero,
$I_{tip}(t)=0$, whereas at the contacts 1 and 2 the potentials are fixed,
$\Delta V_1(t)=\Delta V_2(t)=0$. We measure all voltages relative to the potential
at contact 1 (freedom of the choice of gauge) so that $V_1=0$. Solving the
system of equations (\ref{volfluc1}) for $\Delta V_{tip}(t)$ gives
\begin{equation}
\Delta V_{tip}(t)=-\frac{1}{G_{31}+G_{32}}\Delta I_{tip}(t)
\end{equation}
and the fluctuation spectrum
\begin{equation}
\langle (\Delta V_{tip})^2\rangle=\left(\frac{1}{G_{31}+G_{32}}\right)^2
\langle (\Delta I_{tip})^2\rangle\, . \label{volflucspec}
\end{equation}

Eqs.\ (\ref{cond}), (\ref{shotnoise}) and (\ref{volflucspec})
are our starting points and we apply them
to systems 
consisting of a conductor with two (or more) massive
contacts and one or two weakly coupled contacts as depicted in Figs.\ 1 and 2.
Our plan is
to start with the scattering matrix of the entire system (sample and tip) and
expand this scattering matrix in powers of the coupling strength $|t|$ of the
tip to the conductor. In this way we get equations which contain the scattering
matrices of the separated systems, one describing scattering only in the sample
and one describing the scattering in the tip.

Here, we use the Hamiltonian formulation to express the scattering matrix in
terms of the Green's function of the mesoscopic sample, Eq.\ (\ref{scatgreen}).
Representing the
scattering matrix in terms of Green's functions is a comfortable way to
identify the (non-diagonal) density operators, Eqs.\ (\ref{nondiaginj})
and (\ref{nondiagem}), in the expressions for the
conductances and the current-correlation spectra.

\section{The single tip configuration}

We consider a system consisting of a mesoscopic conductor connected to
$N$ electron reservoirs and which has one additional weakly coupling
contact, the tunneling tip (see e.\ g.\ Fig.\ 1 where $N=2$). The coupling strength between
the tip and the conductor is $|t|$ and the coupling is local at
a point $x$ on the surface of the conductor.

\subsection{Average current at the tip}

The transmission probability at an energy
$E$ for an electron incoming from a massive contact $\alpha$ of the sample
being transmitted into the tip has been found
to be proportional to the injectivity of the contact at the coupling
point $x$ of the tip\cite{gramespacher97},
\begin{equation}
T_{tip,\alpha}=4\pi^2\nu_{tip}|t|^2\nu(x,\alpha)\, .\label{injtip}
\end{equation}
The transmission probability for an electron incoming from the tip being
scattered into a massive contact $\alpha$ is proportional
to its emissivity\cite{gramespacher97},
\begin{equation}
T_{\alpha,tip}=4\pi^2\nu(\alpha,x)|t|^2\nu_{tip}\, .\label{emtip}
\end{equation}
Due to the symmetry of injectivity and emissivity, Eq.\ (\ref{injsym}),
these transmission probabilities manifestly
obey the Onsager-Casimir symmetry, $T_{tip,\alpha}(B)=T_{\alpha,tip}(-B)$,
where $B$ is the magnetic field.
Using these energy resolved transmission probabilities
in Eq.\ (\ref{cond}) we can express
the average current flowing into the tip as
\begin{equation}
\langle I_{tip}\rangle=\frac{e}{h}\int dE T_{ts}(x)\{ f_{tip}(E)-f_{eff}(x)\}
\label{aglg}
\end{equation}
with the two-probe tip-to-sample transmission $T_{ts}(x)=4\pi^2\nu_{tip}
|t|^2\nu(x)$ and the {\em effective local distribution function}
\begin{equation}
f_{eff}(x)=\sum_{\alpha=1}^N\frac{\nu(x,\alpha)}{\nu(x)}f_\alpha(E)\, .
\label{effdis}
\end{equation}
This expression gives the local non-equilibrium distribution of charge carriers
at the
point $x$ inside the conductor. Its energy dependence comes from the Fermi functions
and from a possible energy dependence of the L(P)DOS.

Eq.\ (\ref{aglg}) has the form of the current
in a two probe system, one probe being the tip, where the electron distribution
is described by the Fermi function $f_{tip}(E)$ and the other probe where the
electron distribution is given by the effective distribution function
$f_{eff}(x)$. This effective distribution function does not account for any
energy relaxation of the charge carriers inside the conductor. We assume
that electron-electron and electron-phonon interactions can be neglected
for the system in consideration and therefore the energy of the electrons
is conserved. However, the distribution function does
contain via the L(P)DOS the quantum mechanical phase coherence of the electron
wave function throughout the system. Our effective
distribution can be used to describe the electron distribution in
phase-coherent diffusive conductors, if energy relaxation and dephasing can be
neglected. To describe transport and noise in diffusive conductors one can
also use the semi-classical Boltzmann-equation approach
(see e.\ g.\ \cite{sukhorukov98}).
There, one introduces a
distribution function which does not contain the quantum-mechanical
phase-coherence but where energy relaxation processes can be modeled quite easily.
However, the distribution function of this semi-classical approach can not be
used for conductors where
phase-coherence is essential.

At zero temperature we can replace the Fermi functions in
Eq.\ (\ref{aglg}) by step functions and get in linear response to the applied
potentials
\begin{equation}
\langle I_{tip}\rangle=G(x)\{ V_{tip}-V_{eff}(x)\}\, ,\label{cglg}
\label{avcurtiplin}
\end{equation}
where the conductance $G(x)=(e^2/h)T_{ts}(x)$
has to be taken at the Fermi energy and
\begin{equation}
V_{eff}(x)=\sum_\alpha\frac{\nu(x,\alpha)}{\nu(x)}V_\alpha\, .
\label{veffglg}
\end{equation}
The same formula for the average current is also true for the case of arbitrary
temperature provided that the L(P)DOS, $\nu(x,\alpha)$, are independent of
the energy in an energy interval $\Delta E\approx kT$ around the Fermi-energy.

A particularly interesting setup is, when the tip is used as a voltage probe,
i.\ e.\ we demand that on the average there is no net current flowing into the tip,
$\langle I_{tip}\rangle =0$.
Similar experiments, also called scanning tunneling potentiometry, have
initially been performed by Muralt and Pohl\cite{muralt}
and were later continued and refined by several
groups\cite{kirtley,briner,ramaswamy}. 
From Eq. (\ref{avcurtiplin}) we find that at zero temperature
the voltage one has to apply at the tip
to achieve the zero-current condition
is exactly the effective voltage $V_{eff}(x)$ defined in
Eq.\ (\ref{veffglg}). The measured effective potential $V_{eff}(x)$ should
not be confused with the actual electrostatic potential $U(x)$ inside
the conductor.
The injectivities $\nu(x,\alpha)$ and the LDOS $\nu(x)$
are determined by the equilibrium
electrostatic
potential $U_{eq}(x)$ in the sample\cite{buttikerjphys} and, therefore,
also the measured effective
potential
$V_{eff}(x)$ depends on the electrostatic potential. However, there is
no direct relation between the measured potential
and the actual electrostatic potential in the sample.
\subsection{Current fluctuations at the tip}
We proceed by investigating the fluctuation-spectrum of the current at the tip.
From Eq.\ (\ref{shotnoise}) we get to the lowest order in the coupling parameter
$|t|$,
\begin{eqnarray}
\langle (\Delta I_{tip})^2\rangle & = & 2\int dE G(x)
\left[ f_{eff}(x)\{1-f_{tip}(E)\}\right. \nonumber \\
& + & \left. f_{tip}(E)\{ 1-f_{eff}(x)\}\right]
\label{arbtipfluc}
\end{eqnarray}
with the two-terminal conductance $G(x)$ and the effective distribution
function $f_{eff}(x)$ as defined in Eq.\ (\ref{effdis}).
The fluctuations are therefore, as was the average current, determined by the
effective distribution function.
If we adjust the potential at the tip $V_{tip}$ such that the
average current at the tip vanishes, we get for the fluctuations
\begin{equation}
\langle (\Delta I_{tip})^2\rangle=4\int dEG(x)\{1-f_{tip}(E)\}
f_{eff}(x)\, .\label{tipfluc}
\end{equation}

In Eq.\ (\ref{arbtipfluc})
the integral over energy extends from the bottom of the conduction
band to infinity. At a temperature $T$ and applied potential differences
$\Delta V$, the relevant contribution to the current fluctuations comes from
the integration over an energy range of about $\Delta E\approx\mbox{max}(e\Delta V,
kT)$ around the Fermi-energy. If the LPDOS are nearly independent of energy in
this energy range, we can evaluate the integral over products of Fermi functions
and get for a potential $V_{tip}$
at the tip and potentials $V_\alpha$  at the massive
contacts
\begin{eqnarray}
\langle (\Delta I_{tip})^2\rangle & = & 2eG(x)\sum_{\alpha=1}^N|V_\alpha-V_{tip}|\nonumber \\
& \times & \frac{\nu(x,\alpha)}{\nu(x)}
\coth\left(\frac{e|V_\alpha-V_{tip}|}{2kT}\right)\, .
\end{eqnarray}
If we consider the case of a measurement on a wire with two contacts and choose
$V_{tip}=V_{eff}$ such that on average there is no current flowing into the tip, we
get
\begin{eqnarray}
\langle (\Delta I_{tip})^2\rangle & = & 2eG(x)\Delta V\frac{\nu(x,1)}{\nu(x)}\left(
1-\frac{\nu(x,1)}{\nu(x)}\right)\nonumber \\
& \times & \sum_{\alpha=1}^2\coth\left(\frac{\nu(x,\alpha)}
{\nu(x)}\frac{e\Delta V}{2kT}\right)
\label{fluctemp}
\end{eqnarray}
with $\Delta V=V_1-V_2$. In the limit $e\Delta V\ll kT$ this leads to
\begin{eqnarray}
& &\langle (\Delta I_{tip})^2\rangle \approx 4G(x)kT\nonumber \\
& & \quad +\frac{1}{3}eG(x)\Delta V
\frac{e\Delta V}{kT}\frac{\nu(x,1)}{\nu(x)}\left( 1-\frac{\nu(x,1)}{\nu(x)}
\right)\, ,
\end{eqnarray}
where we neglected corrections of order $(e\Delta V/kT)^2$. In this case the
current fluctuations are due to thermal Johnson-Nyquist noise and a small
correction which depends on the applied bias $\Delta V$.

As a next step we restrict ourselves to the case of zero temperature and
sufficiently small differences in the applied potentials $V_\alpha$
so that we are in the linear response regime. We are then dealing with pure
shot-noise which is completely determined by the properties of the system
(the scattering matrix) at the Fermi energy.
For arbitrary potentials $V_\alpha$ (though always close to
the equilibrium value) we get from Eq.\ (\ref{arbtipfluc}) for the current
fluctuations at the tip
\begin{equation}
\langle (\Delta I_{tip})^2\rangle=2eG(x)\sum_\alpha\frac{\nu(x,\alpha)}
{\nu(x)}|V_\alpha
-V_{tip}|\, . \label{fluccond}
\end{equation}
The conductance, i.\ e.\ the densities of states and the coupling element $t$
contained in it, have to be taken at the
Fermi energy.
This result shows that the fluctuations in the tip are just the addition
of the fluctuations proportional to the conductances between the tip
and the two massive contacts of the wire.
This is not surprising, since, as is well known, the fluctuations of the
current at a
tunneling
contact between two reservoirs are proportional to its
conductance\cite{BdJ}.

Eq.\ (\ref{fluccond}) is valid for arbitrary voltage configurations. Let us now
choose the
potential of the tip such that on average there is no net current flowing into
the tip, i.\ e.\ we have to choose $V_{tip}=V_{eff}(x)$ according to
Eq.\ (\ref{veffglg}). Let us assume that
the applied potentials at the sample are arranged in a way that
$V_\alpha<V_\beta$ for $\alpha>\beta$ and let $n$ be such, that
$V_\alpha>V_{eff}$ for $\alpha\le n$ and $V_\alpha<V_{eff}$ for
$\alpha\ge n+1$.
The fluctuations at
the tip can then be written in the form
\begin{eqnarray}
\langle(\Delta I_{tip})^2\rangle & = & 4eG(x)\sum_{\alpha=1}^n
\frac{\nu(x,\alpha)}{\nu(x)}\{ V_\alpha-V_{eff}(x)\}\nonumber\\
 & = & 4eG(x)\sum_{\alpha=n+1}^N\frac{\nu(x,\alpha)}
{\nu(x)}\{ V_{eff}(x)-V_\alpha\}\, .
\end{eqnarray}
For the case of measurements on a two terminal conductor as shown in
Fig.\ \ref{eintip},
this formula reduces to
\begin{equation}
\langle (\Delta I_{tip})^2\rangle =4eG(x)\Delta V\frac{\nu(x,1)}{\nu(x)}\left(
1-\frac{\nu(x,1)}{\nu(x)}\right)\, .\label{tipnoise}
\end{equation}
with $\Delta V=V_1-V_2$. This shows that at zero temperature 
${\nu(x,1)}/{\nu(x)}$ plays the role of the non-equilibrium 
distribution function. 
\subsection{Voltage fluctuations at the tip}
In the previous section we discussed the fluctuation spectrum of the current
at the tip while we assumed that the potential at the tip is fixed and
independent of time. Let us now investigate the experimental setup shown in
Fig.\ \ref{volflucfig}, where the current at the tip is zero and we measure
the fluctuation spectrum of the voltage using an infinite impedance voltmeter.
If currents and voltages are related by the linear response formula,
Eq.\ (\ref{volfluc1}), the voltage fluctuation spectrum is directly related
to the current fluctuation spectrum, Eq.\ (\ref{volflucspec}). For the case
of zero temperature, we can use the conductances from Eqs.\ (\ref{injtip}) and
(\ref{emtip}) to get the fluctuation spectrum
\begin{equation}
\langle (\Delta V_{tip})^2\rangle = 4eR(x)\Delta V\frac{\nu(x,1)}{\nu(x)}
\left(1-\frac{\nu(x,1)}{\nu(x)}\right) 
\label{tipvolflucspec}
\end{equation}
with $R(x)=G(x)^{-1}$.
Results for the voltage and current fluctuations at finite temperature and in
linear response to the applied potentials are presented in Appendix B.
Next we will illustrate the main
results of the previous section on some examples.
\subsection{Examples}
The most simple example is
a perfect ballistic conductor with one propagating
channel. The local density of states as well as the injectivities are
then independent of position. The injectivities from the left and right contacts
are $\nu_0=1/hv$ and the LDOS is $2\nu_0$. At zero temperature, this gives the
position independent effective voltage $V_{eff}=(V_1+V_2)/2$ and from
Eq.\ (\ref{tipnoise})
the fluctuation spectrum
\begin{equation}
\langle (\Delta I_{tip})^2\rangle=
2eG_0\Delta V\frac{1}{2}\, ,\label{diperfbal}
\end{equation}
with $G_0=(e^2/h) 4\pi^2\nu_{tip}|t|^22\nu_0$ and $\Delta V=V_1-V_2$.
Note that a perfect conducting two terminal conductor shows no fluctuations of
the currents at its contacts.
The presence of the tip introduces shot noise into the system because in the
presence of the tip electrons entering the system from lets say contact 1 have
now the possibility to go either to contact 2 (what they do most of the time)
or to enter the tip (what they do with a probability proportional to $|t|^2$).
The fluctuations at the tip cause also the current at the massive contacts to
fluctuate. At the massive contacts however, there is a considerable
average current of the order of one, while the fluctuations  are only of the
order of $|t|^2$.

As a next step we introduce scattering in the wire. Let us assume that there
is a scattering region described by a scattering matrix which leads to
the transmission probability $T$ and reflection probability $R=1-T$
for the electrons. To the left of the scattering
region the LDOS and the injectivities
are\cite{gasparian}
\begin{eqnarray}
\nu(x,1) & = & \nu_0 (2-T+2\sqrt{1-T}\cos(2kx+\phi))\, \label{erstglg},\\
\nu(x,2) & = & \nu_0 T\, ,\\
\nu(x) & = & 2\nu_0(1+\sqrt{1-T}\cos(2kx+\phi))\, ,\label{letztglg}
\end{eqnarray}
where $\phi$ is the phase acquired by reflected particles.
Putting these densities into the fluctuation spectrum, Eq.\ (\ref{tipnoise}),
leads to
\begin{equation}
\langle(\Delta I_{tip})^2\rangle=
2eG_0\Delta VT\left(1-\frac{T}{2}\frac{1}{1+\sqrt{R}\cos(2kx+\phi)}
\right)\, .
\end{equation}
As a function of the position $x$ of the tip, the fluctuation spectrum
oscillates with a period of half a Fermi wavelength.
If we average this position dependent spectrum over one period
we get
\begin{equation}
\langle (\Delta I_{tip})^2\rangle_{ave}=2eG_0\Delta VT(1-\sqrt{T}/2)\, .
\label{avebarrfluc}
\end{equation}
Averaging over the whole length of the conductor can mean to really move one
single tip along the wire, always adjusting the electro-chemical potential
such that there is zero average current into the tip and
measuring the fluctuation spectrum. But it could also mean to attach very
many tips (or electron absorbers) all
along the wire, each one with its electro-chemical potential adjusted
such that there is no net current flowing into it and neglecting 
the transmission of electrons from one tip to another ($\propto |t|^4$).
It is interesting to compare Eq.\ (\ref{avebarrfluc}) to the fluctuations measured at
contact 1 of an isolated (no tip present) wire\cite{BdJ},
\begin{equation}
\langle(\Delta I_1)^2\rangle=2e\frac{e^2}{h}\Delta VT(1-T)\, .
\end{equation}
Neglecting the interference of incoming and reflected waves in the
local densities, i.\ e.\ setting $\nu(x,1)=\nu_0(2-T)$ and $\nu(x)=2\nu_0$,
one gets form Eq.\ (\ref{tipnoise}),
$\langle(\Delta I)^2\rangle\propto T(1-T/2)$.

\begin{figure}
\epsfxsize8cm
\epsffile{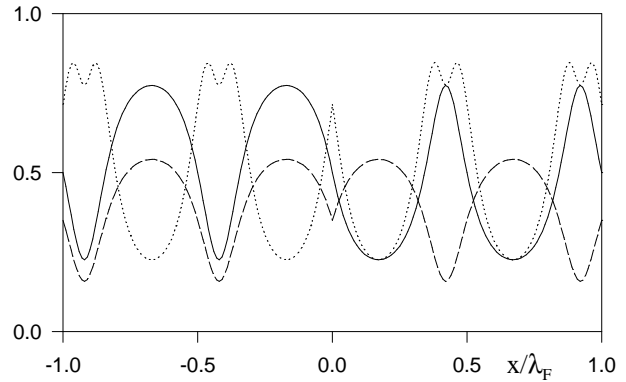}
\vspace{0.3cm}
\caption{Fluctuation spectra and effective voltage measured along a ballistic
wire with a $\delta$ barrier at $x=0$ leading to a transmission probability
$T=0.7$. The solid line is $(V_{eff}-V_2)/\Delta V$, the dashed line gives the
current fluctuations $\langle (\Delta I_{tip})^2\rangle $
in units of $2eG_0\Delta V$,
and the dotted line shows the voltage fluctuations $\langle (\Delta V_{tip})^2
\rangle $ in units of $2eR_0\Delta V$.}
\label{barrfig}
\end{figure}

The voltage fluctuations, Eq.\ (\ref{tipvolflucspec}),
are in the phase-sensitive case, Eqs.\ (\ref{erstglg})-(\ref{letztglg}),
given by
\begin{eqnarray}
\langle (\Delta V_{tip})^2\rangle & = & 2eR_0\Delta V\frac{T}{1+\sqrt{R}
\cos (2kx+\phi)}\nonumber \\
& \times & \left( 1-\frac{T}{2}\frac{1}{1+\sqrt{R}\cos (2kx+\phi)}\right)
\end{eqnarray}
with $R_0=G_0^{-1}$. The current- and voltage-fluctuations together with the
effective potential $V_{eff}(x)$ are shown in Fig.\ \ref{barrfig}.
An interesting system containing very many scatterers
is a metallic diffusive wire of length $L$ and width
$W$ which is at its ends attached to two ideal leads.
The elastic mean free path is $l$. We assume that $l\ll W\ll  L$ so that
the diffusion in the wire can be treated to be effectively one-dimensional.
Furthermore we assume that there is no inelastic scattering inside
the conductor. For a given wire, i.\ e.\ a given disorder configuration,
the exact LPDOS are given in terms of
Green's functions by Eqs.\ (\ref{inj}) and (\ref{em}).
Here, we are only interested in the quantities
averaged over many different disorder configurations.
While the ballistic conductor with one single barrier could serve as a model
to illustrate what is measured in the neighborhood of an impurity, the
ensemble averaged quantities correspond to  the average of many measurements
on a diffusive conductor at different locations over a spatial
range of about an elastic mean free path.
To average expressions
given as products of retarded and advanced Green's functions we use the
diagram technique\cite{altshuler85}.
For the injectivities we have to average the product of retarded and advanced
Green's functions between the coupling point of the tunneling tip and two points
on the surface between the diffusive region and the ideal leads. For the
averaged quantities we get (see appendix A for details)
\begin{equation}
\nu(x,1)=\nu_0(L-x)/L\quad\mbox{and}\quad
\nu(x,2)=\nu_0x/L \label{diffusinj}
\end{equation}
with the two-dimensional density of states $\nu_0=m^\star/2\pi\hbar^2$.

At zero-temperature, the effective voltage measured along the wire gives
averaged over the
ensemble the classical linear voltage drop, $V_{eff}(x)=
V_2+\Delta V(L-x)/L$, and the parabolic behavior of the current
fluctuation spectrum as a function of the tip position,
\begin{equation}
\langle(\Delta I_{tip})^2\rangle=2eG_0\Delta V\frac{x(L-x)}{L^2}\, .
\end{equation}
As in the case of the ballistic conductor with barrier,
we can average the fluctuation spectrum over the hole length of the
diffusive region and get
\begin{equation}
\langle (\Delta I_{tip})^2\rangle = 2eG_0\Delta V\frac{1}{6} \, .
\end{equation}
This is exactly $1/3$ of the fluctuations that would be measured at a tip
probing a perfect ballistic conductor, Eq.\ (\ref{diperfbal}).
It is very well known, that the
fluctuations measured at a contact of a diffusive wire are suppressed by a factor
of $1/3$ with respect to full shot noise (see
e.\ g.\ Refs.\ \cite{nagaev92,beenakker92,sukhorukov98}).
Therefore, it is tempting to say
that the fluctuations at the tip reflect the fluctuations of the current
inside the isolated (without the tip) wire. Nevertheless,
the presence of the tip does change the system
since it offers the electrons another possibility (even though
a very weak one, proportional to $|t|^2$) where to travel. Therefore the
tip introduces
additional fluctuations in the system, as we saw for example when the tip
couples to a perfect ballistic conductor.

\begin{figure}
\epsfxsize8cm
\vspace{0.2cm}
\epsffile{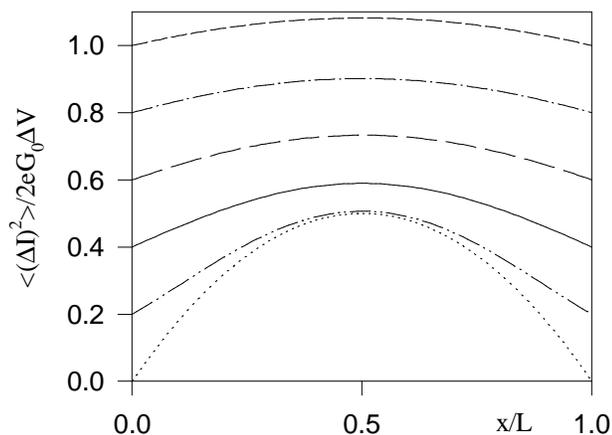}
\vspace{0.3cm}
\caption{The current fluctuation spectrum along a diffusive wire from 0 to $L$
for different temperatures. The temperature range $kT$ is from 0 to $0.5e\Delta
V$ in steps of $0.1e\Delta V$.
Lower temperatures correspond to lower curves.}
\label{crossover}
\end{figure}

\begin{figure}
\epsfxsize8cm
\epsffile{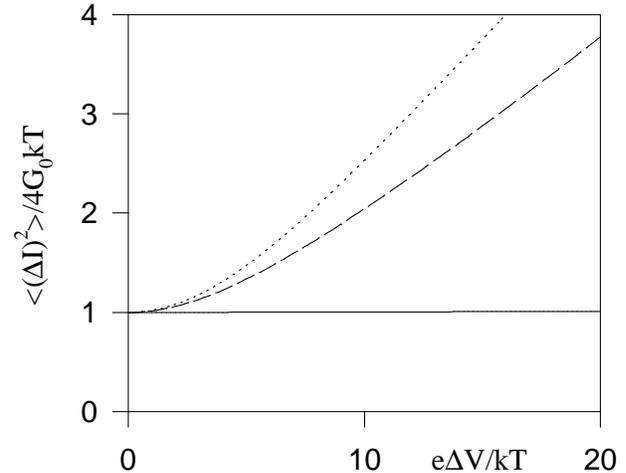}
\vspace{0.3cm}
\caption{Voltage dependence of the current fluctuation spectrum for fixed
temperature. The three curves correspond to different positions of the tip.
The tip is placed at $x=0$ (solid line), $x=L/4$ (dashed line) and $x=L/2$
(dotted line).}
\label{tempdep}
\end{figure}

If we neglect the energy dependence of the injectivities,
Eq.\ (\ref{diffusinj}),
(temperature and
applied bias $\Delta V=V_1-V_2$ sufficiently small) we can use
Eq.\ (\ref{fluctemp}) to illustrate the crossover from the position dependent
shot-noise at zero temperature
to the position independent thermal noise at elevated
temperatures. For the metallic diffusive wire we get
\begin{eqnarray}
\langle ( & \Delta I_{tip} & )^2\rangle  =  2eG_0\Delta V\frac{(L-x)x}{L^2}\nonumber \\
& \times & \left\{ \coth\left(\frac{L-x}{L}\frac{e\Delta V}{2kT}\right)
+ \coth\left(\frac{x}{L}\frac{e\Delta V}{2kT}\right) \right\}\, .
\end{eqnarray}
This crossover is shown
in Fig.\ \ref{crossover}.
In Fig.\ \ref{tempdep}, we plot for
fixed temperature $T$ the voltage dependence of the fluctuation spectrum
if the tip is placed at different positions along the wire.
\newpage
\section{Current correlations at two tunneling probes}
In this Section we make predictions for the cross-correlation
of the currents at two contacts. Recently, two groups succeeded in measuring the
correlation spectrum of the current at two different contacts of a multi-probe
sample\cite{henny98,oliver}. 
We consider a mesoscopic wire with two tips
weakly coupled at points $x$ and $x^\prime$
as shown in figure \ref{zweitips}. For the following discussion, 
we consider the zero
temperature limit and the linear response regime with respect to the applied
potentials.
According to Eq.\ (\ref{shotnoise})
the correlation of the currents
at the two tips $\langle\Delta I_{tip1}\Delta I_{tip2}\rangle$ is
a function of all possible voltage differences $|V_\alpha
-V_\beta|$.
Using the two point density of states, Eq.\ (\ref{nondiaginj}), 
we find 
\end{multicols}
\widetext
\begin{eqnarray} 
\langle\Delta I_{tip1}\Delta I_{tip2}\rangle & = & 2e\frac{e^2}{h}
4\pi^2\nu_{tip1}\nu_{tip2}|t|^4\nonumber\\
& \times & \Big[2Re\{2\pi \nu(x,x^\prime,1)2\pi\nu(x^\prime,x,2)\}|V_1-V_2|
-2Re\{G(x,x^\prime)G(x^\prime,x)\}|V_3-V_4|\nonumber\\
& + & \sum_{\delta=1,2} 2Im\{2\pi \nu(x,x^\prime,\delta)G(x^\prime,x)\}
|V_3-V_\delta|
+\sum_{\delta=1,2}2Im\{2\pi \nu(x^\prime,x,\delta)G(x,x^\prime)\}
|V_4-V_\delta|\Big]\, .\label{allcorr}
\end{eqnarray}
\begin{multicols}{2}
\narrowtext
We now want to illustrate this result for some specific voltage configurations.
One particularly interesting case is the exchange experiment proposed
in \cite{buttiker92a} for arbitrary four terminal conductors. Such an
experiment has been
performed recently by Liu {\em et al.}\cite{liu} on
a ballistic conductor. Theoretical predictions have been made by Blanter
and B\"uttiker\cite{blanter97} and by Sukhorukov and Loss\cite{sukhorukov98}
for metallic diffusive conductors and by van Langen and
B\"uttiker\cite{langen97} for chaotic cavities.
To identify the exchange
contribution in the noise spectrum one performs three successive experiments.
In the first two experiments, called experiment A and B,
current is injected into the system only through
one single contact respectively. In the third experiment, called experiment C,
current is
injected through both contacts simultaneously. The correlation spectrum is always
measured at the same two terminals in all three experiments.
The current injection is achieved by rising the
potential of the respective contact to the elevated value $V_h$ keeping
the other ones at the equilibrium value $V_0$.
In principle, one is free
to choose through which contacts current should be injected and at which two
contacts the correlations should be measured.
In our system we have an obvious
asymmetry between the two massive contacts 1 and 2 of the wire and the two
tunneling contacts 3 and 4. In equation (\ref{allcorr}) we decided to look at the
current correlations at the two tunneling tips. The current correlations at
the two massive contacts will be discussed later. Still we can decide through
which contacts we want to inject the current, either through the massive contacts
or through the tunneling contacts. Experimentally, the first case (contacts
for current
measurement and current injection different) should be
easier to achieve. For both cases we can rewrite Eq.\ (\ref{allcorr}) in the form
\begin{equation}
\langle\Delta I_{tip1}\Delta I_{tip2}\rangle=
-4e\frac{e^2}{h}16\pi^4\nu_{tip1}
\nu_{tip2}|t|^4V S^{m,t}_{A,B,C}\, . \label{currcorrabc}
\end{equation}
Here, the upper index $m$ indicates that the current is injected through the
massive contacts whereas the index $t$ means that current is injected through
the tips. The lower indices distinguish the three experiments and $V=V_h-V_0$.

\subsection{Current injection through the massive contacts}

First we consider the case of current injection through the massive contacts.
Performing the three above mentioned experiments leads to the following voltage
configurations: for experiment A, $V_1=V_h$, for experiment B, $V_2=V_h$
and for experiment C, $V_1=V_2=V_h$. All other potentials are held at
the equilibrium value $V_0$.
We get
\begin{eqnarray}
S_A^m & = & |\nu(x,x^\prime,1)|^2\, ,\label{expam}\\
S_B^m & = & |\nu(x,x^\prime,2)|^2\, ,\label{expbm}\\
S_C^m & = & \frac{1}{4\pi^2}|G(x,x^\prime)-G^\dagger(x,x^\prime)|^2\nonumber\\
& = & |\nu(x,x^\prime,1)+\nu(x,x^\prime,2)|^2\nonumber\\
& = & S_A^m+S_B^m+2\mbox{\rm Re}\{\nu(x,x^\prime,1)\nu(x^\prime,x,2)\}\, .
\label{expcm}
\end{eqnarray}
The current-correlations are for all three experiments determined by the
spatially non-diagonal injectivities, Eq.\ (\ref{nondiaginj}), which
are also given as products of wave functions.
Equations which express the current correlations in terms of
wave functions can be found in \cite{gramespacher98},
Eqs.\ (8)-(11).
It is not surprising that the result for experiment A with current injected
through contact 1 depends only on the (non-diagonal) injectivity of contact 1,
while experiment B with the current injected through contact 2 depends only
on the (non-diagonal) injectivity of contact 2. One sees also at once, that
the result for experiment C is in general not only the addition of experiments
A and B but contains the {\it exchange} term
\begin{eqnarray}
S_X^m & = & S_C^m-S_A^m-S_B^m\nonumber\\
& = & 2\mbox{\rm Re}\{\nu(x,x^\prime,1)\nu(x^\prime,x,2)\}\, .
\label{xchange}
\end{eqnarray}
This exchange term is due to the quantum mechanical {\it indistinguishability} 
of the charge carriers.
In the following we investigate for which systems or under which conditions
this term vanishes or becomes important.
The question if phase-coherence is necessary for the existence of the exchange
term will also be addressed below.

\subsection{Examples}

We investigate Eqs.\ (\ref{expam})-(\ref{expcm})
in more detail for three examples.
The most simple system one can think of is a perfect ballistic
one channel conductor. The two scattering states at the Fermi-energy are
then simple plain waves so that the non-diagonal injectivities at the
Fermi-energy are
\begin{equation}
\nu(x,x^\prime,\alpha)=\frac{1}{hv}e^{ik_\alpha(x-x^\prime)}\, ,
\end{equation}
with $k_1=-k_2=m^\star v/\hbar$ and the Fermi velocity $v=\sqrt{2E_F/m^\star}$.
That means that the correlations in experiments A and B are
independent of the distance $d=x-x^\prime$ of the tips. However, the correlations
of experiment C and therefore the exchange contribution, Eq.\ (\ref{xchange}),
depend on this distance.
They oscillate with the period of half a Fermi-wavelength,
\begin{equation}
S_C^m = 4\frac{1}{h^2 v^2}\cos^2(kd)\, .\label{corperfbal}
\end{equation}
Moving one of the tips over the distance of half a Fermi-wavelength 
and averaging the results, gives the
averaged spectrum
\begin{equation}
\langle S_C^m\rangle=2\frac{1}{h^2 v^2}=S_A+S_B
\end{equation}
which is again independent of the distance between the tips. The exchange term
averages to zero. Moving the tips along the wire means in this case averaging
over the phase of the wave function. Therefore,
for this type of conductors (applies also to perfect ballistic
multi-channel conductors) phase coherence is crucial for the existence of
an exchange term.
A perfect ballistic
(multichannel) conductor exhibits no fluctuations at zero temperature,
and thus the result found above might represent a very particular 
situation. Thus now we introduce scattering in the wire, i.\ e.\ we introduce
a barrier of transmission probability $T$ in the middle of the wire.
This changes the noise properties of the wire in a qualitative way: due to the
possibility of backscattering the current in the massive contact of the wire
fluctuates already without a tip being present. 
Now, we place tip 1 to the left of the barrier and tip 2 to the right of the
barrier. We assume one propagating channel on each side of the
barrier so that the barrier is described by a $2\times 2$ matrix, 
which determines
the scattering states on the two sides. We find 
\begin{eqnarray}
S_A^m & = & 2\nu_0^2 T\left[ 1-T/2-a(2kx-\phi)\right]\, ,\\
S_B^m & = & 2\nu_0^2 T\left[ 1-T/2+a(2kx^\prime+\phi)\right]\, ,\\
S_C^m & = &
2\nu_0^2 T\times 2\cos^2\{k(x-x^\prime)-\phi\} \nonumber \\
& = & 2\nu_0^2 T-2\nu_0^2 T\cos\{2k(x-x^\prime)-2\phi\}\, ,
\end{eqnarray}
with $a(z)=\sqrt{1-T}\sin(z-\phi_a)$.
Here, $\nu_0=1/hv$ is the density of scattering states,
$\phi$ is the phase acquired by an
electron traveling through the barrier and $\phi_a$ is the phase 
which takes into account
a possible asymmetry of the barrier\cite{gasparian}.
The spectrum of experiments A and B in which current is injected only through
one single contact depends only on the position of the tip at that side of
the barrier where current is injected. 
Comparison of the spectrum of experiment C for the pure 
ballistic wire Eq.\ (\ref{corperfbal}),
with that for a wire with a barrier shows 
that these spectra 
differ only in that the spectrum of the wire with a scatterer 
is multiplied by the transmission probability $T$ and in that it 
depends on $\phi$, the phase acquired by
transmitted electrons. Again, we find a non-vanishing exchange
term $S_X^m=S_C^m-S_A^m-S_B^m$.
Moving the tips on both sides of the barrier over a Fermi wavelength does not
cause the exchange term to vanish, but leads to
\begin{equation}
S_X^m=-2\nu_0^2 T(1-T)\, .
\end{equation}
Thus elastic scattering has established 
a correlation in the exchange term which does not vanish upon 
averaging. 

It is an interesting question whether an exchange term exists also for
measurements on diffusive conductors or not. Starting from exact quantum
mechanical expressions for the correlation spectrum and performing a disorder
average, Blanter and B\"uttiker\cite{blanter97} found a non-vanishing
exchange term for cross shaped diffusive conductors.
An exchange term for diffusive four-terminal conductors of arbitrary
shape was found by 
Sukhorukov and Loss\cite{sukhorukov98} using a semi-classical Maxwell-Boltzmann
equation approach which does not contain the quantum mechanical phase
coherence of the system.
In our approach, we start with the quantum mechanical expressions for the
non-diagonal injectivities, Eq.\ (\ref{nondiaginj})
and average these quantities over many different disorder
configurations. We assume the conductor to be a long and narrow strip as 
discussed in Section IV,
and, similarly to Ref. \cite{blanter97} 
use the diagram technique to average products
of Green's functions.
Performing the averages leads to the following expressions for the noise
spectra\cite{gramespacher98} (details see appendix A)
\begin{eqnarray}
S_A^m & = & \frac{S_C}{2}\frac{(L-x)^2+(L-x^\prime)^2+p(x,x^\prime)}{L^2}\, ,\\
S_B^m & = & \frac{S_C}{2}\frac{x^2+(x^\prime)^2+p(x,x^\prime)}{L^2}\, , \\
S_C^m & = & \frac{(m^\star)^2}{(\pi\hbar)^2N}\frac{L}{l}\frac{x(L-x^\prime)}
{L^2}\, ,
\end{eqnarray}
where $p(x,x^\prime)=1/3[(x-x^\prime)^2-2x^\prime(L-x)]$. From these results we
can extract the relative strength of the exchange term $S_X^m$ to be
\begin{equation}
\frac{S_X^m}{S_C^m}=
\frac{1}{L^2}\left[ x(L-x)+x^\prime(L-x^\prime)-p(x,x^\prime)
\right]\, .
\end{equation}
The exchange term always has the same sign as the spectra $S_A^m$ and
$S_B^m$, i.\ e.\ 
it enhances the correlation spectrum $S_C^m$ over the pure addition
$S_A^m+S_B^m$.
An enhancement of the current correlations due to the exchange term was also
predicted for a chaotic cavity with four tunneling contacts\cite{langen97}.
To illustrate the exchange term further, we assume a specific configuration
of the two tips: we place the two tips symmetrically around the center $L/2$
of the wire. One is placed a distance $d/2$ to the left of the center, the other
one the same distance $d/2$ to the right. The strength of the exchange
term as a function of the distance $d$ between the tips is then
\begin{equation}
\frac{S_X^m}{S_C^m}=
\frac{1}{3}\left( 2+\frac{d}{L}-2\left(\frac{d}{L}\right)^2
\right)\, .
\end{equation}
This  function reaches its maximum not when the tips are closest
(a limit where our approximations for the disorder average are not anymore
valid),
but at the finite distance $d=L/4$.
It's maximal value is $(S_X^m/S_C^m)_{max}=17/24$.
At first sight, it might seem quite surprising to have the maximal correlations
when the tips are separated by $d=L/4$. This can be understood if one considers 
that the strength of the correlations is determined by scattering between
all four contacts and, therefore, not only the distance in between the tips counts,
but also the distances from the coupling points of the tips to the massive
contacts of the wire.
Moving the tips away from the center of the wire increases the distance
in between them, but decreases the distances to the massive contacts. The
correlations are then determined by an interplay of contributions from the
differing types of possible electron paths. This example again demonstrates 
that in the presence of elastic scattering, the exchange contribution
survives ensemble averaging. This is consistent with the results 
of Refs.\ \cite{blanter97,sukhorukov98}. 

\begin{figure}
\epsfxsize8cm
\epsffile{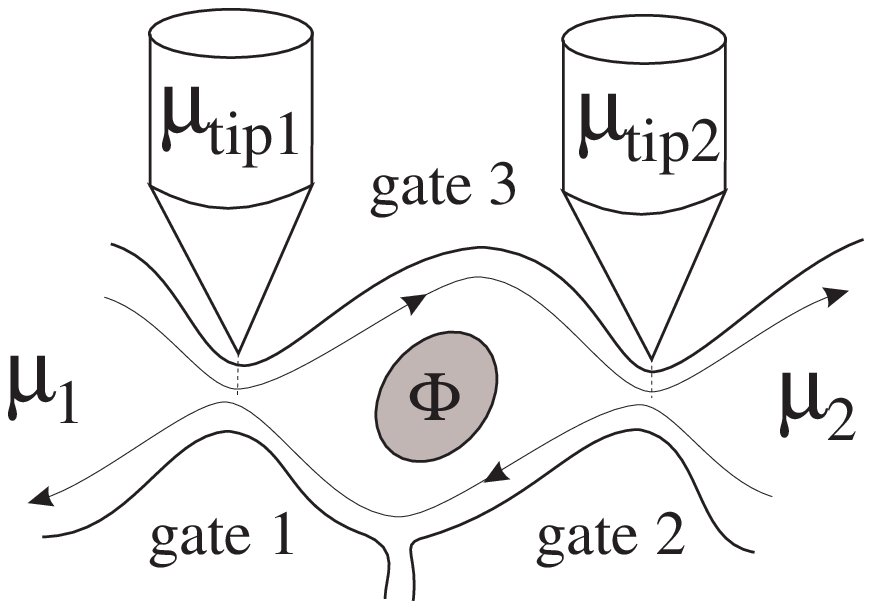}
\vspace{0.3cm}
\caption{Mesoscopic ring in the quantum Hall regime with one propagating edge
channel. An additional magnetic flux penetrates the center of the ring which
is not accessible to the electrons. Two tunneling tips are placed
at the center of the point contacts which connect the ring to two electron
reservoirs.}
\label{ring}
\end{figure}

Let us consider as a last example
a system consisting of a quantum dot in the quantum Hall regime,
to which two leads are
attached via quantum point contacts, Fig.\ \ref{ring}.
A very similar geometry was investigated in Ref. \cite{buttiker92b}. 
We re-consider this example since the two-point injectivity
Eq. (\ref{nondiaginj}) provides a particularly clear formulation 
and also to use this opportunity to correct an algebraic mistake 
in one of the results of Ref. \cite{buttiker92b}. 
The sample is penetrated by a quantizing magnetic field which leads to the formation
of edge channels. The voltages at the gates forming the two point contacts are
chosen
such that there is exactly one propagating edge channel which is perfectly
transmitted through the sample whereas all other edge channels are completely
reflected at the point contacts. In addition to the strong magnetic field there is an
additional field present only in the center of the dot. The additional field is
characterized by its flux $\Phi$ through the dot. Since there is no backscattering
at all of electrons in the propagating edge channel, the transmission probability
of the system is independent of the flux $\Phi$.
For the same reason transmission from one tip to the other or to the massive
contacts is also independent
of $\Phi$.
Without backscattering there is no closed electron path encircling the flux.
Now we place two tunneling
tips in the middle of the two point contacts 1 and 2. There, the tips should couple
equally well to the left going and to the right going edge channel.
We are interested if the correlation of the currents at the two tips
depends on $\Phi$. To answer this question we only need to know the scattering wave
functions at the two coupling points. Let us denote the amplitude of the 
scattering state incoming
from the left contact at the left point contact by $\psi_1(1)$ and the one incoming
from the right contact at the right point contact by $\psi_2(2)$.
The electron state
$\psi_1$ acquires now on its way from the left to the right point contact
an additional phase
$\phi_1$ due to its propagation and the presence of the background quantizing
magnetic field. In addition, it's phase is changed by $\theta/2$ due to the
flux $\Phi$. Therefore, we have $\psi_1(2)=\psi_1(1)e^{i\phi_1}e^{i\theta/2}$.
Similarly, we have $\psi_2(1)=\psi_2(2)e^{i\phi_2}e^{i\theta/2}$. As before,
$\phi_2$ is the phase acquired due to propagation and the presence of the
quantizing field and $\theta/2$ is due to the flux $\Phi$.
For all closed paths encircling the flux one must have
\begin{equation}
\oint_S {\bf A}d{\bf s}=2\pi\frac{\Phi}{\Phi_0}=\theta\, .
\end{equation}
We chose a gauge such that the phase $\theta$ is divided into equal parts 
on the upper and the lower half circle along the edge of the dot.
Putting these wave functions into the expression for the current correlations
at the tips, Eq.\ (\ref{expcm}), yields
\begin{equation}
S_C^m\propto 2+2\cos\{(\phi_1+\phi_2)+\theta \}
\end{equation}
i.\ e.\ the exchange term is $S_X^m=2\cos\{(\phi_1+\phi_2)+\theta \}$. (we
have used
$|\psi_1(1)|^2=|\psi_2(2)|^2=1$). We see that the correlation spectrum
in fact depends periodically on the flux and the period is $\Phi_0$.
The measurement of the correlation spectrum thus allows to get information about
the flux $\Phi_0$.
This result corrects Eq. (15) of Ref. \cite{buttiker92b}
where the periodicity of the
correlation spectrum was found to be only $\Phi_0/2$.
We remark that the exchange term depends on the phases 
$\phi_1$ and $\phi_2$ in a similar simple way as the exchange 
term of the pure ballistic wire. Again, moving the tips by the distance 
of a Fermi wavelength will lead to a vanishing exchange term. 
Furthermore this example shows that a cross-correlation can be sensitive to 
an Aharonov-Bohm flux even for a conductor (which in the absence of the 
tips) exhibits no Aharonov-Bohm effect.
However, the situation discussed here and in Ref.\ \cite{buttiker92b}
does not conclusively show that Aharonov-Bohm effects in second order 
correlations are possible even if there is no second order 
Aharonov-Bohm effect. If the conductance is measured in the presence of the 
two tips, then the weak scattering caused by the tips, which must after all
couple to both edge states, leads to an Aharonov-Bohm effect, which is of 
the same magnitude (fourth order in the tunneling amplitudes) as the 
fourth order interference effect given by the current-current 
correlation. 

\subsection{Current injection through the tips}

We now consider slightly modified arrangements: instead of injecting the current
through the massive contacts, we inject the current through the tips and measure
simultaneously the correlations of the currents at the tips. The voltage
configurations for the three experiments of this type are then for experiment
A, $V_3=V_h$, for experiment B, $V_4=V_h$ and for experiment C, $V_3=
V_4=V_h$. All other potentials are held at the equilibrium value
$V_0$. The correlation spectrum for experiment C is the same as the spectrum
of experiment C with the current injected through the massive contacts since
the spectrum depends only on the absolute value of voltage differences and not
on the sign. Experiments A and B are, however, different from the respective
experiments with current injection through the massive contacts.
The quantities $S_{A,B,C}^t$ which have to be used in Eq.\ (\ref{currcorrabc})
are
\begin{eqnarray}
S_A^t & = & \frac{1}{4\pi^2}|G(x^\prime,x)|^2\, ,\label{expat} \\
S_B^t & = & \frac{1}{4\pi^2}|G(x,x^\prime)|^2\, ,\label{expbt}\\
S_C^t & = & \frac{1}{4\pi^2}|G(x,x^\prime)-G^\dagger(x,x^\prime)|^2
\nonumber\\
& = & S_A^t+S_B^t-\frac{1}{2\pi^2}\mbox{\rm Re}\{G(x,x^\prime)G(x^\prime,x)\}\, ,
\label{expct}
\end{eqnarray}
Since the potentials of both massive contacts are always held at the same
potential, the
equilibrium potential $V_0$, the correlation spectra do
not show any dependence on the (non-diagonal) injectivities of these two
contacts separately. 
The wire acts as an effective one terminal conductor
and all that enters in Eqs.\ (\ref{expat})-(\ref{expct})
is the Green's function of the wire
representing the total (non-diagonal) density of states of the wire.
But as in the experiments discussed in the previous section
an exchange term appears in general.
To investigate this exchange term further we evaluate it for the example systems
used before.

For a ballistic wire the result is qualitatively similar
to the one found by current injection through the tunneling contacts.
A qualitative change occurs for the wire with a barrier
and in the case
of a metallic diffusive wire.
Let us first consider the ballistic wire with the barrier.
In contrast to the experiments where current is injected through the massive
contacts, the averaged exchange term does vanish when current is injected through
the tips.
Averaging means to move both tips over distances longer
than a Fermi wavelength and average the measured spectra.
For a metallic diffusive conductor it is easily seen that the exchange term
vanishes. The average over disorder of a product of two retarded Green's
functions
is exponentially small. This is in remarkable contrast to the behavior of
the exchange term in the  experiments with current injection through the
massive contacts.
It is due to the fact, that the spectrum of experiments A and B changed
while experiment C is the same as for current injection through massive
contacts.

We can draw the following conclusions from this Section:
For all the situations investigated here, we could identify an 
exchange contribution to the cross-correlation. 
In the case of a pure ballistic wire, the exchange contribution 
is a purely quantum mechanical effect which vanishes when averaging 
is performed (by moving the tip and averaging the results). 
As soon as some elastic scattering is present, as in the wire 
with a barrier, or in a metallic diffusive wire, the exchange term,
in addition to a purely quantum mechanical contribution, also 
contains a "classical" contribution which survives ensemble 
averaging. This situation is thus reminiscent of the conductance 
of a mesoscopic sample which consists of a classical (Drude like)
conductance and a small quantum mechanical sample specific 
contribution known as universal conductance fluctuation.  

\section{current correlation at the massive contacts}
Until now we were only interested in the correlation of the currents at the 
tunneling tips. We saw that in the case of current injection through the 
massive contacts the correlations depend on non-diagonal partial densities of
states, namely the non-diagonal injectivities of the massive contacts.
There is still another
partial density of states, the emissivities, which did not yet appear in the
expressions for the correlation spectra. Emissivity and injectivity are
related to each other by the symmetry relation, Eq.\ (\ref{injsym}). The correlations
of the currents at the tips were determined by the transport properties of
electrons injected through the massive contacts and transmitted to the tips.
Therefore only the non-diagonal injectivities of the massive contacts appeared
in the equations for the current correlations at the tips.
If one investigates the correlation spectrum at
the massive contacts one expects that it depends on the non-diagonal
emissivities of these contacts.
Clearly, if we also inject the current through the massive contacts, the
correlation of the current in the massive contacts is to first order only
determined by the wire with its two contacts and the presence of the two tips
does not play a role at all. In this case, the correlation/fluctuation spectra
are just the ones known for two probe conductors\cite{BdJ}.

Consider the case, when current is injected through the tips. We investigate 
the experiments A: $V_3=V_h$, B: $V_4=V_h$ and C: $V_3=V_4=V_h$.
All other potentials are, as before, kept at $V_0$. The correlations
can then be written in the form
\begin{equation}
\langle\Delta I_1\Delta I_2\rangle=-2e\frac{e^2}{h}16\pi^4|t|^4eV S_{A,B,C}
\end{equation}
with
\begin{eqnarray}
S_A & = & \nu(1,x)\nu(2,x)\nu_{tip1}^2\, ,\\
S_B & = & \nu(1,x^\prime)\nu(2,x^\prime)\nu_{tip2}^2\, ,\\
S_C & = & \nu(1,x)\nu(2,x)\nu_{tip1}^2+
\nu(1,x^\prime) \nu(2,x^\prime)\nu_{tip2}^2\nonumber \\
& & +2\mbox{\rm Re}\{\nu(1,x,x^\prime)\nu(2,x^\prime,x)\}\nu_{tip1}\nu_{tip2}\\
& = & S_A+S_B+2\nu_{tip1}\nu_{tip2}\mbox{\rm Re}
\{\nu(1,x,x^\prime)\nu(2,x^\prime,x)\}
\, . 
\end{eqnarray}
The expressions for experiments A and B are products of the transmission
probabilities from tip 1 resp.\ tip 2 into the two massive contacts of the
wire, e.\ g.\ the transmission probability from tip 1 into contact 1 of the
wire is $T_{1,tip1}=4\pi^2\nu(1,x)|t|^2\nu_{tip1}$ according to
Eq.\ (\ref{emtip}).
The two spectra where current is only injected into the system through one
single contact do not at all depend on the presence of the second tip.
They depend only on the local emissivities of the massive contacts at the
coupling point of the tip through which the current is injected.
The correlation spectrum of experiment C where current is injected through
both tips is sensitive to the non-diagonal
emissivities of the massive contacts.
In fact, the exchange contribution is 
\begin{eqnarray}
S_X=2\nu_{tip1}\nu_{tip2}
\mbox{\rm Re}\{ \nu(1,x,x^\prime)\nu(2,x^\prime,x)\}
\end{eqnarray}
This result again demonstrates the key role played by the 
the two point injectivity in cross-correlation spectra.  

\section{Discussion}

We have shown that the current fluctuation and correlation spectra measured
at tunneling contacts on multi-probe conductors are related to local partial
densities of states and to spatially non-diagonal (two point) 
densities of states. The
general expressions are illustrated for various examples, like perfect
ballistic conductors, ballistic conductors with a barrier, metallic
diffusive wires and mesoscopic rings in a magnetic field.

In particular, we found that the current fluctuations at a single tunneling tip
are determined by an {\it effective local distribution function}
$f_{eff}(x)$.
This distribution function is given in terms of local partial densities of
states, the injectivities of the contacts of the sample, $f_{eff}(x)=
\sum_\alpha (\nu(x,\alpha)/\nu(x))f_\alpha(E)$.
It gives the local non-equilibrium distribution of charge carriers in a conductor.
In the semi-classical Boltzmann-equation approach
one relates the current fluctuations to local distribution functions. These
distribution functions are solutions to the Boltzmann-equation with proper
boundary conditions. They do not contain the quantum mechanical phase
coherence of an electron state entering through contact $\alpha$ and traveling
to the point $x$ in the conductor, whereas this information is included
via the densities of states in our distribution function $f_{eff}(x)$.
Our discussion bridges therefore at least to some extend the gap between 
quantum mechanical discussions of shot noise and purely classical treatments
of current fluctuations. 
The effects of the phase coherence on the
fluctuation spectrum is illustrated for measurements on a ballistic conductor
with a barrier. This example is also useful to get a qualitative
impression on how the noise spectrum looks like in the neighborhood of an 
impurity.
We evaluate the general formula for the 
fluctuations at the tip also for the case of measurements on a metallic
diffusive wire in the ensemble average.

The second part of this work treats the current correlations in two tunneling
contacts. The correlations are determined by newly defined spatially non-diagonal
and non-local densities of states. We used 
the exchange experiment\cite{buttiker92a} to investigate the magnitude of the
exchange term in a four terminal configuration containing two tunneling tips.
If current is injected through the massive contacts of the sample, the
correlation spectrum at the tips is given by the  spatially non-diagonal
injectivities $\nu(x,x^\prime,\alpha)$. If current is injected through the
tips, the correlation spectrum at the massive contacts is given by the
non-diagonal emissivities $\nu(\alpha,x,x^\prime)$.
An exchange term with a magnitude of the order of the total
correlations was found for ballistic conductors and ballistic conductors
with a barrier. The correlations are always negative while the exchange term
can have either sign, depending on the positions of the tips. This can lead to a
complete suppression of the correlations for certain tip positions.
Even for the case of measurements on metallic diffusive
conductors an exchange term exists, and it's magnitude can be as high as 70 \%
of the total correlations. In the average over the disorder configurations, the
exchange term is always negative and therefore enhances the correlations.
For the example of a mesoscopic ring penetrated by a magnetic flux we showed,
that the current correlations measured in the tips can show a flux
dependence even though the conductances through the ring do not depend on the
flux.

Clearly, the experiments proposed here, if carried out, would permit 
an unprecedented, detailed microscopic view of shot noise in mesoscopic 
conductors. 

This work was supported by the Swiss National Science Foundation.
\begin{appendix}
\section{Ensemble averages for diffusive wires}
We consider a two-dimensional metallic diffusive wire of length $L$ and
width $W$ with $L\gg W$. The elastic mean free path is $l\ll W$. Then
the diffusion can be treated to be effectively one-dimensional.
The diffusive wire is at its ends connected via a coupling matrix
$\Gamma_\alpha$ to two semi-infinite ideal leads.
\subsection{Ensemble averaged injectivity}
We are looking for the disorder average of the injectivity of contact $\alpha$
at a point $r=(x,y)$ inside the diffusive region, Eq.\ (\ref{inj}).
We have to find the average of
\begin{equation}
\nu(r,\alpha)=\int_{S_\alpha} dy_1dy_2G(r,r_1)\Gamma_\alpha(y_1,y_2)
G^\dagger(r_2,r)\, .\label{injint}
\end{equation}
Here, the integrals are over the surface between contact $\alpha$ and the
diffusive region.
The coupling matrix $\Gamma_\alpha(y,y^\prime)$ is independent of the disorder
configuration inside the wire. The disorder average of Eq.\ (\ref{injint}) is
then
\begin{eqnarray}
\langle\nu(r,\alpha)\rangle & = & \int_{S_\alpha}dy_1dy_2\Gamma_\alpha(y_1,y_2)
\nonumber\\
& \times & 
\int dr_a S(r,r_a)\langle G(r_a,r_1)\rangle\langle
G^\dagger(r_2,r_a)\rangle\, .
\end{eqnarray}
The integral over the intermediate point $r_a$ is over the entire diffusive
region. The propagator
\begin{equation}
S(r,r^\prime)=\frac{1}{D\tau WL}\left\{{x^\prime(L-x)\qquad x>x^\prime\atop
x(L-x^\prime)\qquad x<x^\prime}\right.\label{props}
\end{equation}
with the diffusion coefficient $D=v_Fl/2$ and the elastic lifetime $\tau=l/v_F$ 
describes the diffusion from the point $r$ to $r^\prime$. 
In particular, this propagation is independent of the $y$ coordinate provided
that $|x-x^\prime|\gg l$. The exponentially
decaying averaged Green's functions can be approximated as
\begin{equation}
\langle G(r,r^\prime)\rangle=-\frac{im^\star}{\hbar p_F}\exp\left[
\left(ip_F-\frac{1}{2l}\right)|x-x^\prime|
\right]\delta(y-y^\prime)\, .\label{aveg}
\end{equation}
Performing the integrals and using
$\int dy_\alpha \Gamma_\alpha(y_\alpha,y_\alpha)
=v_F N_\alpha/4\pi$ ($N_\alpha=k_FW$ is
the number of open channels in contact $\alpha$) then gives
\begin{equation}
\langle\nu(r,1)\rangle=\nu_0\frac{L-x}{L}\, ,
\end{equation}
and
\begin{equation}
\langle\nu(r,2)\rangle=\nu_0\frac{x}{L}\, .
\end{equation}
Here, we used the two dimensional density of states $\nu_0=m^\star/2\pi\hbar^2$.
The injectivities are linearly decaying, respectively, growing as functions of
the position $x$ along the wire. They are independent of the transverse
coordinate $y$.
\subsection{Ensemble averaged non-diagonal injectivity}
\begin{figure}
\epsfxsize5.5cm
\centerline{\epsffile{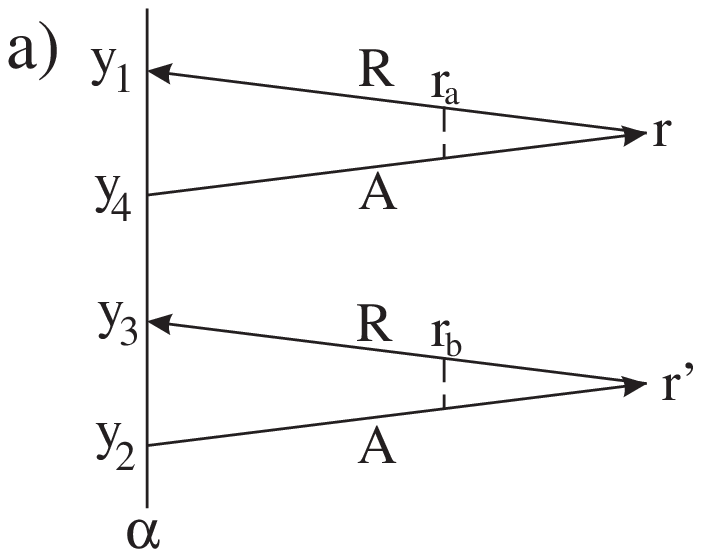}}
\vspace{0.2cm}
\epsfxsize5.5cm
\centerline{\epsffile{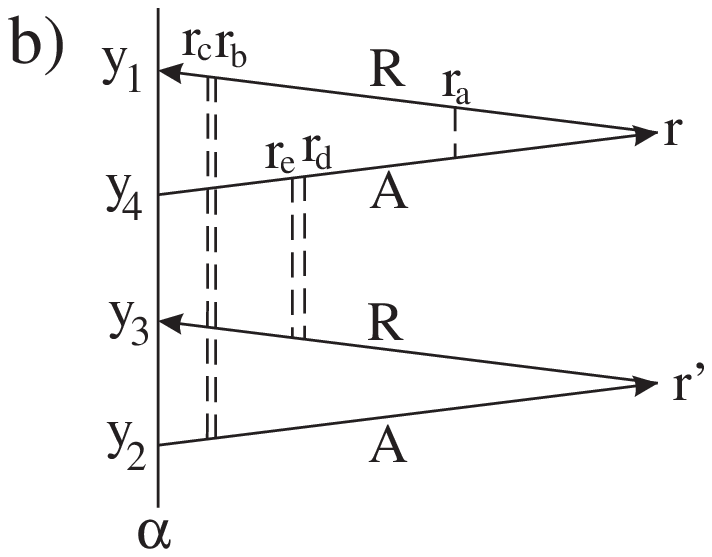}}
\vspace{0.2cm}
\epsfxsize5.5cm
\centerline{\epsffile{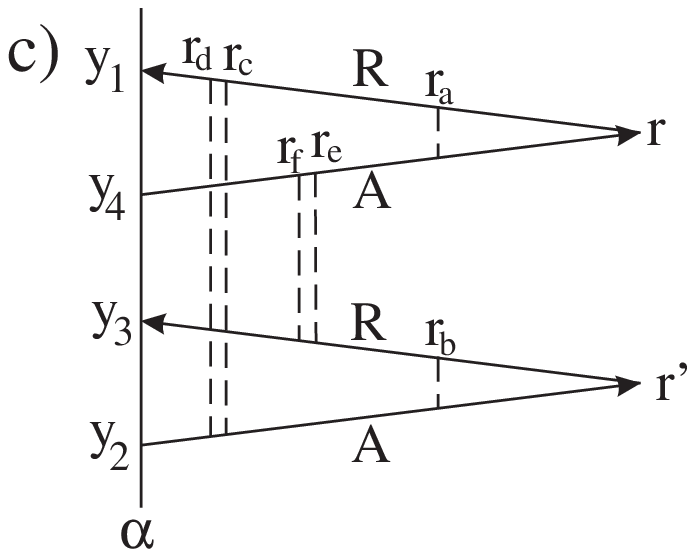}}
\caption{Diagrams for the average of the four Greens functions using (a) two 
diffusions, (b) three diffusons and (c) four diffusons. A single dashed
line indicates the propagation with the propagator $S(r,r^\prime)$ and
two neighboring dashed lines indicate propagation with $P(r,r^\prime)$.}
\label{diagramme}
\end{figure}

In Section V we found that the current correlations were proportional
to absolute squares of non-diagonal injectivities,
\begin{eqnarray}
|\nu(r,r^\prime,\alpha)|^2 & = & \int_{S_\alpha}dy_1dy_2dy_3dy_4
\Gamma_\alpha(y_1,y_2)\Gamma_\alpha(y_3,y_4)\nonumber\\
& \times &
G(r_,r_1)G^\dagger(r_2,r^\prime)G(r^\prime,r_3)G^\dagger(r_4,r)\, .
\end{eqnarray}
Now, we are interested in the average of this quantity over many different
disordered wires.
Again, the $\Gamma$'s are independent of the impurity configuration inside the
wire, so that it remains to find the average of the product of four
Green's functions.
The averaged quantity has contributions from diagrams with two, three and
four diffusion propagators, as shown in figure \ref{diagramme}.
It is interesting to compare these diagrams for the two-point injectivity 
with the ones given by Blanter and one of the authors \cite{blanter97}
which apply in a discussion of the shot noise at the contacts 
of metallic diffusive conductors . 
It turns out that
diagrams with two and three diffusons are small as $l/L$, respectively
$(l/L)^2$, compared to the diagram with four diffusons and are therefore
neglected. From the diagram with four diffusion propagators we get
\begin{eqnarray}
& &
\langle|\nu(r,r^\prime,\alpha)|^2\rangle =
\int_{S_\alpha}dy_1dy_2dy_3dy_4\Gamma_\alpha(y_1,y_2)\Gamma_\alpha(y_3,y_4)
 \nonumber\\
& & \quad\times 
\int dr_adr_bdr_cdr_ddr_edr_f S(r,r_a)S(r^\prime,r_b) \nonumber \\
& & \quad\times P(r_c,r_d)P(r_e,r_f)
\langle G(r_d,r_1)\rangle\langle G^\dagger(r_2,r_d)\rangle \nonumber \\
& & \quad\times \langle G(r_f,r_3)
\rangle\langle G^\dagger(r_4,r_f)\rangle
 F(r_b,r_c,r_a,r_e)\, . 
\end{eqnarray}
Here, the averaged Green's functions and the propagator $S(r,r^\prime)$ are
given by Eqs.\ (\ref{props}) and (\ref{aveg}), and
\begin{equation}
P(r,r^\prime)=\frac{1}{\hbar^3 m^\star D\tau^2 WL}\left\{ {
x(L-x^\prime)\qquad x<x^\prime \atop x^\prime (L-x)\qquad x>x^\prime}\right.
\, .
\end{equation}
$F(r_1,r_2,r_3,r_4)$ is the short-ranged Hikami box\cite{hikami} 
and in Fourier space is given by\cite{blanter97} 
\begin{eqnarray}
& & F(q_1,q_2,q_3,q_4)= -m^\star(\tau/\hbar)^5
v_F^5(2\pi)^2\delta(q_1+q_2+q_3+q_4)  \nonumber \\
& & \quad\times [2(q_1q_3+q_2q_4)+(q_1+q_3)(q_2+q_4)]\, .
\end{eqnarray}
Performing all the integrals then gives the result
\begin{eqnarray}
\langle |\nu(r,r^\prime,1)|^2
\rangle
 & = & 2\left(\frac{m^\star}{2\pi\hbar^2}\right)^2\frac{1}{k_Fl}\frac{1}{WL}
\frac{x(L-x^\prime)}{L^2}p(x,x^\prime)
\nonumber\\
& = & \frac{\nu_0^2}{g}\frac{x(L-x^\prime)}{L^4}
p(x,x^\prime)
\end{eqnarray}
with the abbreviation $p(x,x^\prime)=(L-x)^2+(L-x^\prime)^2+\frac{1}{3}
(x-x^\prime)^2-\frac{2}{3}x^\prime (L-x)$.
In the last step we used the Drude conductance $g=k_FWl/2L$.
The results for $\langle |\nu(r,r^\prime,2)|^2\rangle$ and
$\langle\nu(r,r^\prime,1)\nu(r^\prime,r,2)\rangle$ are obtained using the
same procedure.
\section{Finite temperature linear response results}
For the configuration of figure \ref{eintip}, Eq.\ (\ref{aglg}) gives the average
current at the tip at fixed temperature and for given potentials
$\mu_\alpha$ at the massive contacts and $\mu_{tip}$ at the tip. In this section
we are interested in the case of finite temperature $T$ and small applied
bias such that $kT\gg \Delta \mu$. In this limit we can approximate the Fermi
functions $f_\alpha(E)$ in the reservoirs of the massive contact $\alpha$ of
the sample
with the help of the Fermi function in the reservoir of the tip,
\begin{equation}
f_\alpha(E)\approx f_{tip}(E)-\frac{\partial f_{tip}}{\partial E}(\mu_\alpha
-\mu_{tip})\, . \label{fexp}
\end{equation}
Using this expansion in (\ref{aglg}) we get
\begin{equation}
\langle I_{tip}\rangle = \frac{e}{h}\sum_\alpha\int dET_{ts}(x)
\left(-\frac{\partial f}{\partial E}\right)\frac{\nu(x,\alpha)}{\nu(x)}
(\mu_\alpha - \mu_{tip})
\end{equation}
with the Fermi function $f(E)$ describing the distribution of electrons in
the reservoir of the tip held at a potential $\mu_{tip}$. If we want to use the
STM as a voltage probe we can easily solve the equation $\langle I_{tip}\rangle
=0$ for $\mu_{tip}$ and find
\begin{equation}
\mu_{tip}=\frac{\sum_\alpha \int dE T_{ts}(x)\left(-\frac{\partial f}{\partial E}
\right)\frac{\nu(x,\alpha)}{\nu(x)}\mu_\alpha}{\int dE T_{ts}(x)\left(
-\frac{\partial f}{\partial E}\right)}\, .
\label{vmess}
\end{equation}
If one can take the fraction $\nu(x,\alpha)/\nu(x)$ to be (nearly) independent
of energy in an interval of size $kT$ around the Fermi energy \cite{note2},
equation (\ref{vmess}) reduces to the result valid at zero temperature,
Eq.\ (\ref{veffglg}). 

To find the finite temperature linear-response current fluctuation spectrum
at the tip we have to insert the expansion (\ref{fexp}) into
Eq.\ (\ref{tipfluc}). This gives
\begin{eqnarray}
\langle (\Delta I_{tip})^2\rangle & = & 4\int dE G(x)\left(-\frac{\partial f}
{\partial E}\right)\nonumber \\
& \times & \left\{ kT +f(E)\sum_\alpha\frac{\nu(x,\alpha)}{\nu(x)}
(\mu_\alpha-\mu_{tip})\right\}\, ,
\end{eqnarray}
where $\mu_{tip}$ is adjusted according to Eq.\ (\ref{vmess}) such that the
average current at the tip vanishes. The current fluctuations are the 
addition of pure thermal, Johnson-Nyquist noise, $\langle
(\Delta I_{tip})^2\rangle_{therm}=4G_{eff}(x)kT$ with the effective conductance
$G_{eff}(x)=\int dE G(x)(-\partial f/\partial E)$ and an excess noise proportional
to the applied bias. Using an infinite impedance external circuit to measure
the voltage at the tip, Eq.\ (\ref{volflucspec}),
gives the voltage fluctuation spectrum
\begin{eqnarray}
\langle (\Delta V_{tip})^2\rangle & = & 4 R_{eff}(x)kT\nonumber\\
& + & 4R_{eff}(x)^2\int dE G(x)\left( -\frac{\partial f}{\partial E}\right)f(E)
\nonumber \\
& \times &
\sum_\alpha \frac{\nu(x,\alpha)}{\nu(x)}(\mu_\alpha-\mu_{tip})
\end{eqnarray}
with the effective resistance $R_{eff}(x)=[G_{eff}(x)]^{-1}$.

\end{appendix}

\end{multicols}

\begin{references}
%
\bibitem{engquist}
H.-L.\ Engquist and P.\ W.\ Anderson, Phys.\ Rev.\ B {\bf 24}, 1151 (1981).
%
\bibitem{landauerres}
R.\ Landauer, Phil.\ Mag.\ {\bf 21}, 863 (1970);
IBM J.\ Res.\ Develop.\ {\bf 1}, 223 (1957).
%
\bibitem{buttiker86}
M. B\"{u}ttiker, Phys. Rev. Lett. 57, 1761 (1986);
IBM J. Res. Developm. 32, 317 (1988)
%
\bibitem{buttiker89}
M.\ B\"uttiker, Phys.\ Rev.\ B {\bf 40}, 3409 (1989).
%
\bibitem{gramespacher97}
T.\ Gramespacher and M.\ B\"uttiker, Phys.\ Rev.\ B {\bf 56}, 13026 (1997).
%
\bibitem{imry} Using Fermi Golden Rule arguments weak coupling 
contacts sensitive to currents (but not amplitudes) have been treated by
Y.\ Imry, in {\it Directions in Condensed Matter Physics},
edited by G.\ Grinstein and G.\ Mazenko, (World Scientific Singapore, 1986).
p.\ 101. 
%
\bibitem{binnig82}
G.\ Binnig and H.\ Rohrer, Helv.\ Phys.\ Acta {\bf 55}, 726 (1982);
G.\ Binnig {\em et al.}, Phys.\ Rev.\ Lett.\ {\bf 49}, 57 (1982).
%
\bibitem{wiesendanger}
{\em Scanning Tunneling Microscopy I,II,III}, Springer Series in Surface Sciences 20,
28, 29, Ed.\ by R.\ Wiesendanger and H.-J.\ G\"untherodt (Springer,
Heidelberg, 1992).
%
\bibitem{avouris95}
Ph.\ Avouris, I.-W.\ Lyo, and Y.\ Hasegawa, IBM J.\ Res.\ Dev.\ {\bf 39},
603 (1995), and other articles in the same issue.
%
\bibitem{crommie}
M.\ F.\ Crommie, C.\ P.\ Lutz, and D.\ M.\ Eigler, Science {\bf 262}, 218
(1993).
%
\bibitem{dekker}
L.\ C.\ Venema, J.\ W.\ G.\ Wild\"oer, J.\ W.\ Janssen,
S.\ J.\ Tans, H.\ L.\ J.\ Temminck Tuinstra, L.\ P.\ Kouwenhoven,
and C.\ Dekker, cond-mat/9811317.
%
\bibitem{tersoff}
J.\ Tersoff and D.\ R.\ Hamann, Phys.\ Rev.\ B {\bf 31}, 805 (1985).
%
\bibitem{bardeen}
J.\ Bardeen, Phys.\ Rev.\ Lett.\ {\bf 6}, 57 (1961).
%
\bibitem{bracher}
C.\ Bracher, M.\ Riza, and M.\ Kleber, Phys.\ Rev.\ B {\bf 56}, 7704 (1997).
%
\bibitem{pothier}
H.\ Pothier, S.\ Gu\'eron, N.\ O.\ Birge, D.\ Esteve, and M.\ H.\ Devoret,
Phys.\ Rev.\ Lett.\ {\bf 79}, 3490 (1997).
%
\bibitem{buttikerjphys}
M.\ B\"uttiker, J.\ Phys.\ Condens.\ Matter {\bf 5}, 9361 (1993).
%
\bibitem{christen} M.\ B\"uttiker and T.\ Christen, in:
{\it Quantum Transport in Semiconductor Submicron Structures},
Ed.\ by B.\ Kramer, NATO ASI Series, Vol.\ {\bf 326} (Kluwer,
Dordrecht, 1996), p.\ 263.
%
\bibitem{christen96}
T.\ Christen and M.\ B\"uttiker, Europhys.\ Lett.\ {\bf 35}, 523 (1996).
%
\bibitem{ma98} Z.\ S.\ Ma, J.\ Wang, and H.\ Guo, Phys.\ Rev.\ B {\bf 57},
9108 (1998).    
%
\bibitem{buttiker92a}
M.\ B\"uttiker, Phys.\ Rev.\ B {\bf 46}, 12485 (1992).
%
\bibitem{BdJ}
M.~J.~M.~de~Jong and C.~W.~J.~Beenakker, in:
{\em Mesoscopic Electron Transport}, Ed.\ by L.~L.~Sohn,
L.\ P.\ Kouwenhoven, and G.\ Sch\"on, NATO
ASI Series E, Vol. {\bf 345} (Kluwer, Dordrecht, 1997),
p.\ 225.
%
\bibitem{glattli}
L.\ Saminadayar, D.\ C.\ Glattli, Y.\ Jin, and B.\ Etienne,
Phys.\ Rev.\ Lett.\ {\bf 79}, 2526 (1997).
%
\bibitem{picci}
R.\ de-Picciotto, M.\ Heiblum,
V.\ Umansky, G.\ Bunin, and D.\ Mahalu,
Nature (London) {\bf 389}, 162 (1997).
%
\bibitem{brom}
H.\ E.\ van den Brom and J.\ M.\ van Ruitenbeek, cond-mat/9810276.
%
\bibitem{birk95} 
H.\ Birk, M.\ J.\ M.\ de Jong, and C.\ Sch\"onenberger, Phys.\ Rev.\ Lett.\ 
{\bf 75}, 1610 (1995).
%
\bibitem{buttiker90}
M.\ B\"{u}ttiker, Phys.\ Rev.\ Lett.\ {\bf 65}, 2901 (1990).
%
\bibitem{buttiker92b}
M.\ B\"uttiker,
Phys.\ Rev.\ Lett.\ {\bf 68}, 843 (1992).
%
\bibitem{thrl}
Th.\ Martin and R.\ Landauer, Phys.\ Rev.\ B {\bf 45},
1742 (1992).
%
\bibitem{blanter97}
Ya.\ M.\ Blanter and M.\ B\"uttiker, Phys.\ Rev.\ B {\bf 56}, 2127 (1997).
%
\bibitem{sukhorukov98}
E.\ V.\ Sukhorukov and D.\ Loss, Phys.\ Rev.\ Lett.\ {\bf 80}, 4959 (1998);
cond-mat/9809239.
%
\bibitem{langen97}
S.\ A.\ van Langen and M.\ B\"uttiker, Phys.\ Rev.\ B {\bf 56}, R1680 (1997).
%
\bibitem{anantram}
M. P. Anantram and S. Datta, Phys. Rev. B{\bf 53}, 16390 (1996);
S. Datta and P. F. Bagwell, and M. P. Anantram, Phys. Low-Dim. Struct. 
{\bf 3}, 1 (1996). 
%
\bibitem{henny98}
M.\ Henny, S.\ Oberholzer, C.\ Strunk, and C.\ Sch\"onenberger, (unpublished);
M.\ Henny, Thesis (University of Basel, 1998).
%
\bibitem{oliver}
W.\ D.\ Oliver, J.\ Kim, R.\ C.\ Liu, and Y.\ Yamamoto, (unpublished).
%
\bibitem{buttiker91}
M. B\"{u}ttiker, Physica B{\bf 175}, 199 (1991).
%
\bibitem{liu} 
R.~C.\ Liu, B.\ Odom, Y.\ Yamamoto, and S.\ Tarucha,
Nature {\bf 391}, 263 (1998).
%
\bibitem{gramespacher98}
T.\ Gramespacher and M.\ B\"uttiker, Phys.\ Rev.\ Lett.\ {\bf 81}, 2763 (1998).
%
\bibitem{iida90}
S.\ Iida, H.\ A.\ Weidenm\"uller, and J.\ Zuk, Phys.\ Rev.\ Lett.\ {\bf 64},
583 (1990); Ann.\ Phys.\ (N.Y.) {\bf 200}, 219 (1990).
%
\bibitem{note1} Note that the total charge consists of the injected charge 
and the screening charge due to the long range Coulomb interaction. See 
Ref. \onlinecite{buttikerjphys,christen}. Here the charge generated 
by an external potential variation, keeping the internal potential fixed, 
is of interest. 
%
\bibitem{dattabook}
S.\ Datta, {\em Electronic Transport in Mesoscopic Conductors} (Cambridge
University Press, Cambridge, England, 1995).
%
\bibitem{muralt}
P.\ Muralt and D.\ W.\ Pohl, Appl.\ Phys.\ Lett.\ {\bf 48}, 514 (1986).
%
\bibitem{kirtley}
J.\ R.\ Kirtley, S.\ Washburn, and M.\ J.\ Brady, Phys.\ Rev.\ Lett.\ {\bf
60}, 1546 (1988).
%
\bibitem{briner}
B.\ G.\ Briner, R.\ M.\ Feenstra, T.\ P.\ Chin, and
J.\ M.\ Woodall, Phys.\ Rev.\ B {\bf 54}, R5283 (1996).
%
\bibitem{ramaswamy}
G.\ Ramaswamy and A.\ K.\ Raychaudhuri,
cond-mat/9812167.
%
\bibitem{gasparian}
V.\ Gasparian, T.\ Christen, and M.\ B\"uttiker, Phys.\ Rev.\ A {\bf 54},
4022 (1996).
%
\bibitem{altshuler85}
B.\ L.\ Altshuler and A.\ G.\ Aronov, in: {\em Electron-electron Interactions
in Disordered Systems}, Ed.\ by A.\ L.\ Efros and M.\ Pollak (North-Holland,
Amsterdam, 1985), p.1.
%
\bibitem{nagaev92} K.~E.~Nagaev, Phys.\ Lett.\ A {\bf 169}, 103 (1992).
%
\bibitem{beenakker92}
C.\ W.\ J.\ Beenakker and M.\ B\"uttiker, Phys.\ Rev.\ B {\bf 46}, 1889 (1992).
%
\bibitem{note2}
This is for instance the case if the energy intervall $kT$ is smaller
than the Thouless energy.
%
\bibitem{hikami}
S.\ Hikami, Phys.\ Rev.\ B {\bf 24}, 2671 (1981). 
%
\end{references}
\end{document}